# Galactic Cosmic-Ray Anisotropies: *Voyager 1* in the Local Interstellar Medium

J. S. Rankin[1,2], E. C. Stone[2], A. C. Cummings[2], D. J. McComas[1], N. Lal[3], B. C. Heikkila[3]

[1]*Department of Astrophysical Sciences, Princeton University, Princeton, New Jersey, USA*
[2]*California Institute of Technology, Pasadena, California, USA*
[3]*Goddard Space Flight Center, Greenbelt, Maryland, USA*

## Abstract

Since crossing the heliopause on August 25, 2012, *Voyager 1* observed reductions in galactic cosmic ray count rates caused by a time-varying depletion of particles with pitch angles near 90°, while intensities of particles with other pitch angles remain unchanged. Between late 2012 and mid-2017, three large-scale events occurred, lasting from ~100 to ~630 days. Omnidirectional and directional high-energy data from *Voyager 1*'s Cosmic Ray Subsystem are used to report cosmic ray intensity variations. Omnidirectional ($\gtrsim$ 20 MeV) proton-dominated measurements show up to a 3.8% intensity reduction. Bi-directional ($\gtrsim$ 70 MeV) proton-dominated measurements taken from various spacecraft orientations provide insight about the depletion region's spatial properties. We characterize the anisotropy as a "notch" in an otherwise uniform pitch-angle distribution of varying depth and width centered about 90° in pitch angle space. The notch averages 22° wide and 15% deep – signifying a depletion region that is broad and shallow. There are indications that the anisotropy is formed by a combination of magnetic trapping and cooling downstream of solar-induced transient disturbances in a region that is also likely influenced by the highly compressed fields near the heliopause.

## 1. Introduction

*Voyager 1*'s crossing of the heliopause on August 25, 2012, was marked by sharp increases in low-energy galactic cosmic rays (GCRs) and corresponding sudden decreases in anomalous cosmic rays, as observed by the Cosmic Ray Subsystem (CRS) and Low Energy Charged Particle (LECP) instruments (Stone et al. 2013; Webber & McDonald 2013; Krimigis et al. 2013). In the wake of Voyager's interstellar arrival, LECP observed an unexpected anisotropy in the GCRs characterized by a clear reduction in >211-MeV proton intensities for particles entering their bi-directional telescope when viewing perpendicular to the magnetic field. Several extended, time-dependent events have continued to occur during *Voyager 1*'s interstellar journey beyond the heliopause.

In addition to the GCR anisotropies, *Voyager 1*'s four working instruments observed several signatures of transient disturbances in the interstellar medium. Burlaga et al. (2013) and Burlaga & Ness (2016) reported several weak, laminar, quasi-perpendicular, subcritical, resistive disturbances observed by the magnetometer. Gurnett et al. (2013) and Gurnett et al. (2015) detail a series of locally-generated electron plasma emissions detected by the Plasma Wave (PWS) instrument. Moreover, they compare the PWS-measured events with GCR disturbances seen by CRS and LECP and describe their relationship as analogous to precursor effects often observed in the "foreshock" region upstream of planetary bow shocks.

Evidence suggests that the transient events and GCR anisotropies may be related. For example, Jokipii & Kóta (2014) and Kóta & Jokipii (2017) showed numerical simulations indicating that a gradual compression, followed by a slow weakening of the magnetic field may account for the pitch angle and time profiles of both transient GCR increases and anisotropic decreases. These authors interpreted the pitch-angle anisotropies to arise from particle trapping and adiabatic cooling behind these weak disturbances.

While *Voyager 1* was making these detailed observations of the particle distributions just beyond the heliopause, the Interstellar Boundary Explorer (IBEX, McComas et al. 2009a) was imaging the 3-D properties and structure of the heliosphere's global interaction with the interstellar medium. In particular, IBEX discovered a "ribbon" of enhanced energetic neutral atom (ENA) emissions associated with the local interstellar magnetic field, which drapes around the heliosphere (McComas et al. 2009b; Schwadron et al. 2009). The ribbon provides the best determination of the external field direction (Funsten et al. 2009, 2013) and magnitude (~0.29 nT) at great distances (>1000 au) (Zirnstein et al. 2016). The draping and compression of this interstellar field around the outside of the heliopause leads to higher magnetic field strengths at *Voyager 1*, consistent with its local observations (Pogorelov et al. 2017) and even higher field strengths closer to the heliopause in the IBEX ribbon directions (McComas et al. 2009a, Pogorelov et al. 2011).

IBEX observations also revealed the importance of the interstellar magnetic field in shaping the global heliosphere. These showed that the interstellar medium's magnetic pressure produces large-scale asymmetries in the heliosphere's global structure, with the largest compression and greatest pressure region in the inner heliosheath, between the termination shock and heliopause, ~20⁰ south and slightly offset toward the port side from the interstellar upwind direction (Schwadron et al. 2014). This offset pressure maximum causes asymmetric plasma flows in the inner heliosheath and naturally explains the unexpected flow directions observed by *Voyager 2* (McComas & Schwadron 2014). IBEX observations and the global asymmetries they expose in the heliopause's shape are also important for understanding the detailed particle distributions observed by *Voyager 1*, as we show in this study.

In the following, we focus on CRS measurements of the GCR anisotropy, presenting additional information about these unusual events through measurements of proton-dominated intensities. We describe CRS telescope modes that are relevant to viewing the anisotropy and report observations for varying spacecraft orientations in Section 2. In Section 3, we model the temporal and spatial behavior of the unexpected pitch angle phenomena and in Section 4, we incorporate the results into three types of simulated response functions for comparison with observations. Finally, in Section 5, we explore magnetic trapping and shock-related adiabatic cooling as possible physical mechanisms for producing the anisotropy.

## 2. Particle Anisotropy Observations

The Cosmic Ray Subsystem's double-ended high-energy telescopes (HETs 1 & 2) have geometry factors and energy ranges appropriate for observing GCR intensities and spectra in the local interstellar medium (LISM). Each telescope is composed of circular energetic particle detectors arranged in a cylindrical stack. Both HETs consist of 7 silicon solid-state detectors (C1 through C4) with annular guard rings (G) that operate as omnidirectional anti-coincidence counters. The end detectors consist of 2 thin detectors on the A-end (A1 & A2) and 2 curved detectors on the B-end (B1 & B2) (Stone et al. 1977). To provide directional measurements for multiple species over various energy ranges, CRS telescopes operate in multiple coincidence modes. Those of relevance to this study include HET 1 & 2's bi-directional penetrating mode (PENH; proton-dominated[1], $\gtrsim$ 70 MeV) and omnidirectional mode (Guards; proton-dominated, $\gtrsim$ 20 MeV).

Figure 1 shows LECP and CRS count rates in the LISM from 2012.5 through 2017. LECP's >211 MeV protons show the anisotropy's signatures in the form of long-duration, time-varying intensity changes, present in Sectors 1 & 5 but not in other sectors. LECP has an advantage for viewing the pitch angle anisotropy because its stepper-motor platform routinely steps through eight viewing directions, as indicated by the circular diagram in Figure 1a.

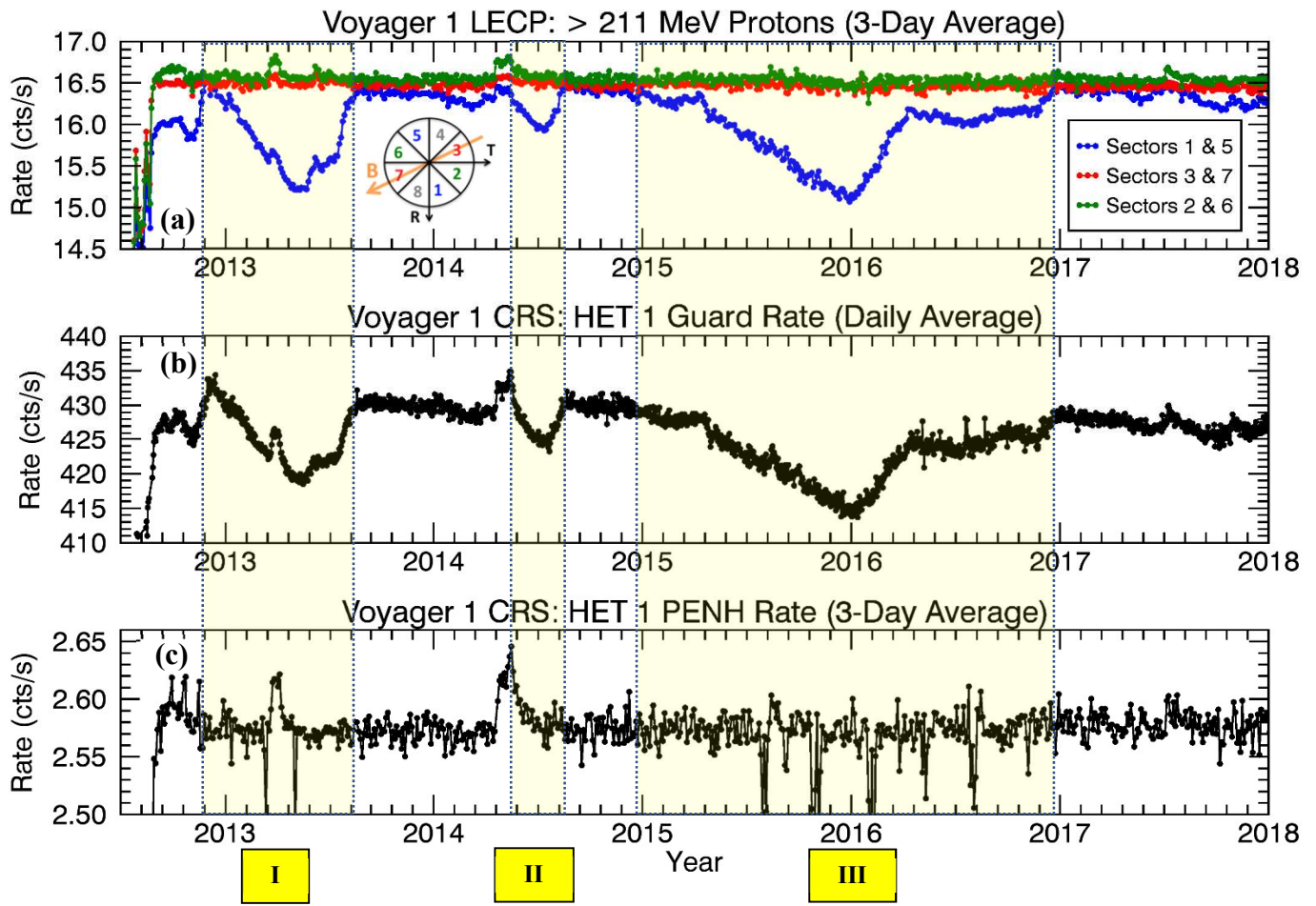

**Figure 1.** LECP and CRS counting rates in the LISM from 2012.5 through 2017. The three largest anisotropy episodes (shaded in yellow) last ~265 (I), ~100 (II), and ~630 (III) days, respectively.

---

[1] In addition to protons, PENH is ~25% electrons and ~5% heavier nuclei ($\gtrsim$ 70 MeV/nuc). See Cummings et al. (2016) for more details on the constituents of CRS rates.

(a) LECP's >211 MeV protons reveal the GCR pitch angle anisotropy. The magnetic field direction lies in Sectors 3 & 7, while Sectors 1 & 5 are approximately perpendicular to the field direction, as illustrated by the circular diagram (background-corrected data is courtesy of Rob Decker and the LECP team; for LECP's non-corrected, publicly-available data, see: http://sd-www.jhuapl.edu/VOYAGER/index.html).
(b) CRS's omnidirectional guard rate (≳ 20 MeV; proton-dominated) from anti-coincidence counters on the HET 1 telescope show similar time dependence to LECP's Sectors 1 & 5.
(c) CRS's bi-directional PENH rate on HET 1 (≳ 70 MeV; proton-dominated) is fairly steady in the LISM, in agreement with LECP's bi-directional rates in Sectors 2 & 6 and 3 & 7. Two types of deviation arise from: (1) shock-related increases (e.g. 2014.35), and (2) decreases observed during 70°-offset spacecraft maneuvers (e.g. 2015.59).

CRS's omnidirectional counters (Figure 1b) continuously monitor the temporal intensity changes without providing pitch angle information. Nevertheless, the omnidirectional guard rates have the highest statistics of all the rates available on CRS (several hundred counts/sec) and show a time-varying intensity response similar to LECP's. Detecting the anisotropy using directional observations presents a greater challenge. CRS's telescopes are body-fixed on the 3-axis stabilized spacecraft and HET 1 & HET 2 fields of view do not typically observe particles with ~90° pitch angles, so their nominal rates are not sensitive to the anisotropy (see for example, HET 1's bi-directional PENH rate in Figure 1c). However, data taken during occasional spacecraft maneuvers provide an opportunity to examine the pitch angle variation of the intensity at specific times.

## 2.1 Magnetometer Roll Maneuvers and Observations

Magnetometer roll maneuvers are performed ~6 times per year for magnetometer calibration purposes. They originally consisted of 10 successive 360° turns about the spacecraft's Earth-pointing axis (approximately $\hat{R}$ in the R, T, N coordinate system)[2], but as of 2017 they are performed with a reduced number of turns because of power limitations. During the 10-roll period (~5.6 hours), CRS telescope fields of view smoothly and continuously rotate 360° every 2,000 s (0.18 °/sec), which translates to an 8.6° angular averaging per point in the highest-resolution data (48-s). "Clock-angle" refers to the angle of the boresight in the N-T plane with the $\hat{N}$-axis as the origin and the angle increasing towards $\hat{T}$. "Roll interval" refers to the set of 10 turns which took place on a particular day (e.g. the 2015-310 interval occurred on day 310 of 2015). Knowing the roll rate, the magnetic field direction, and the clock angle orientation of a telescope's boresight enables the average pitch angle of particles entering the telescope to be determined during each 48-s period throughout a roll maneuver.

HET 1 and HET 2 bi-directional PENH measurements during roll maneuvers confirm that the reduction observed by LECP's Sectors 1 & 5 (Figure 1a) and CRS's omnidirectional rates (Figure 1b) results from a pitch angle anisotropy. Moreover, roll maneuver data provide the clearest measure of the anisotropy's spatial distribution. Figure 2 displays a superposition of HET 1 (Figure 2a) and HET 2 (Figure 2b) rates during 7 rolls where the anisotropy is most prominent (selected intervals are indicated in Tables 3 & 4 of Appendix A). Although the effects of its time-variable magnitude are also present, not only does the anisotropy occur within a region that is perpendicular to the magnetic field – in agreement with LECP's observations – but it is distributed about 90° pitch angle.

---

[2] R, T, N is a spacecraft-centered coordinate system where $\hat{R}$ is the sun-to-spacecraft vector, $\hat{T}$ is the cross product of the sun's rotation vector with $\hat{R}$, and $\hat{N}$ completes the triad of the right-handed system.

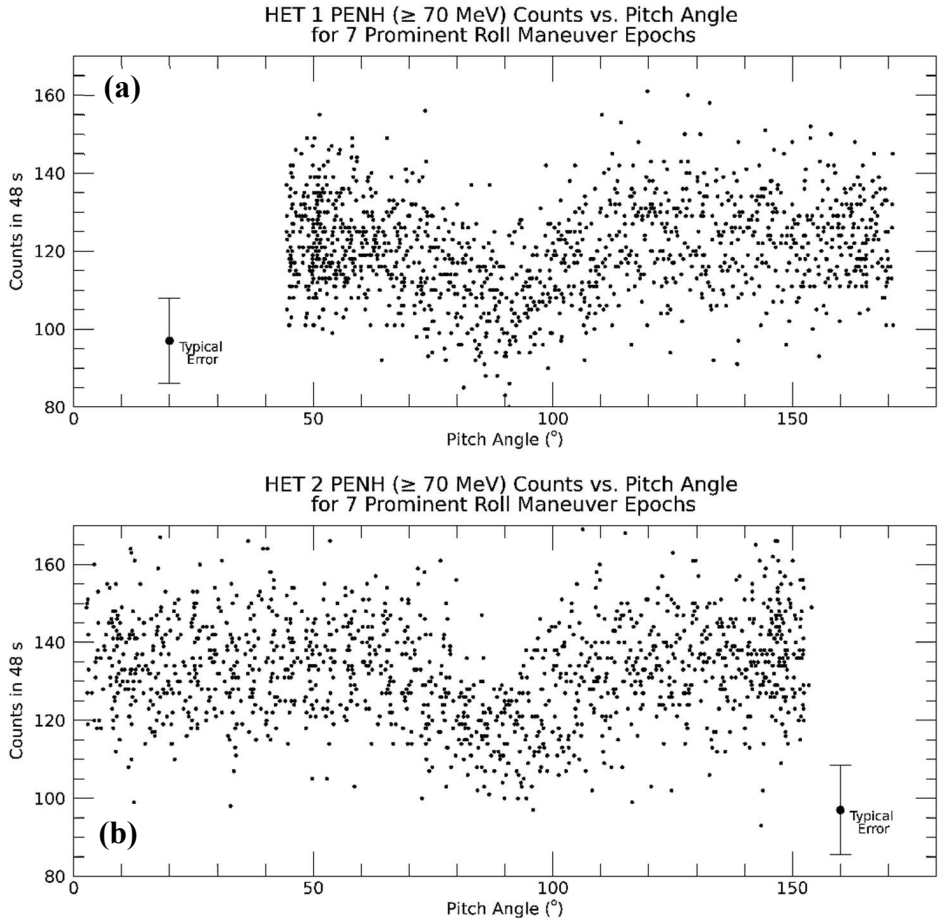

**Figure 2.** Superposition of 7 prominent HET 1 (a) and HET 2 (b) roll maneuver intervals of varying anisotropy magnitudes arranged in pitch angle space (see Appendix A, Tables 3 & 4 for selected intervals).

## 2.2 70°-offset maneuvers and Observations

70°-offset maneuvers were introduced on *Voyager 1* in March 2011 to provide a way for LECP to measure heliosheath plasma flow velocity in the direction not seen in its usual configuration (Decker et al. 2012) and were discontinued in 2017. Like magnetometer roll maneuvers, they require the spacecraft to roll about the $\hat{R}$-axis. However, rather than rolling continuously, the spacecraft rotates to a clock angle offset of 70° and parks for up to 5 hours before returning to its usual orientation. These maneuvers typically occur on consecutive days over a multiple-day period, usually near times of roll maneuvers.

For each offset period, we combine counts from multiple days and normalize to temporally adjacent non-offset values to determine a relative intensity change arising from the pitch-angle anisotropy ($\delta_{70°}$). Table 1 compares average HET 1 & 2 boresight pitch angles for times when the spacecraft is in its usual configuration and to those during 70°-offsets. In Figure 3 we show the average HET 1 PENH rates during the 2015-296 "offset interval", where DOY 296 is the first day of the sequence of 7 maneuvers that took place on days 296 to 312 of 2015 – this is the interval nearest to the 2015-310 roll maneuver. During 70°-offsets, HET 1's field of view overlaps significantly with 90° pitch angle (Table 1), thus enabling these fixed-orientation measurements to complement roll maneuver and omnidirectional observations of the pitch angle anisotropy.

| Telescope | Average Nominal Boresight Pitch Angle | Average 70°-offset Boresight Pitch Angle | Field of View (Full Angle) |
|---|---|---|---|
| HET 1 | 136° ± 3° (A-end)<br>44° ± 3° (B-end) | 77° ± 3° (A-end)<br>103° ± 3° (B-end) | 40° (PENH) |
| HET 2 | 31° ± 4° (A-end)<br>149° ± 4° (B-end) | 69° ± 3° (A-end)<br>111° ± 3° (B-end) | 40° (PENH) |

**Table 1.** A summary of CRS telescope boresight directions in pitch angle space. Note that particles entering a given telescope travel in directions opposite to the telescope's average boresight direction and field of view. Averages were determined using telescope and magnetic field directions from ~2012.65 to 2017.0. The average magnetic field during this period was (0.143, -0.401, 0.179) nT in R, T, N. Uncertainties primarily reflect the small variations in the magnetic field direction.

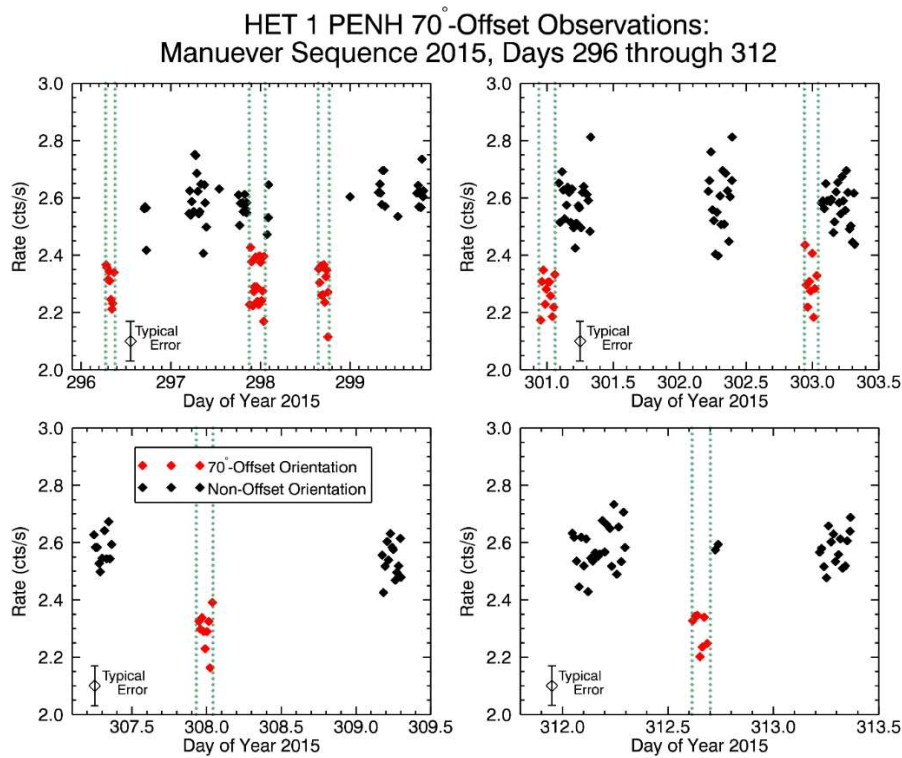

**Figure 3.** HET 1 PENH (≳70 MeV; proton-dominated) 70°-offset observations for a full sequence of maneuvers that took place in 2015, on days 296 to 312. Note that offset maneuvers take place within a subset of time over a period of multiple days, in contrast with roll maneuvers which take place on a single day. The roll maneuver nearest to this 2015-296 offset occurred on day 310 of 2015. HET 1's average 70°-offset boresight pitch angle during this time is 79.3° (A-end). Points are averaged over ~480-s intervals for visualization purposes. The large data gaps show *Voyager 1*'s daily gaps in communication with Earth. The dotted lines denote the times when the spacecraft was fixed in the offset position. The red points mark observations taken while the spacecraft was parked at the 70°-clock-angle offset from its usual position. The black points represent values obtained while the spacecraft was in its nominal orientation. Since HET 1's ~40°-wide field of view includes 90° pitch angle during 70°-offsets (Table 1), it sees a reduction of counts indicative of the anisotropy.

## 2.3 Omnidirectional Observations

We calculate the omnidirectional intensity reduction ($\delta_{omni}$) by comparing observations of each period's daily average to the average rate during 2013.6 to 2014.1 – a time when the pitch angle anisotropy is not prominent (note the steady rates in Figures 1a & 1b). The average isotropic rates used for normalization are 430.01 ± 0.06 counts/sec for HET 1 and 382.38 ± 0.06 counts/sec for HET 2.

The three main episodes of GCR intensity changes caused by the pitch-angle anisotropy (Figure 1b) last on the order of 265 (region I), 100 (region II), and 630 (region III) days and exhibit at most 2.6%, 1.3%, and 3.8% intensity reductions, respectively. A fundamental characteristic of the anisotropy that is supported by CRS's omnidirectional, roll maneuver, and 70°-offset observations is that these long-duration intensity changes arise primarily from the pitch-angle anisotropy, as opposed to effects such as solar modulation, a radial gradient, or diffusive or convective flows.

# 3. Characterizing the Anisotropy

We model the anisotropy by generating particle distributions that are isotropic except for a "notch" centered at 90° pitch angle. Such a notch of missing particles could be either partial or complete, so we base our simulation on two parameters – the notch's width and its depth – and compare the results to the overall reduction in the observed omnidirectional and directional GCR intensities. The actual pitch angle distribution might be more complicated than even a partially-depleted, field-perpendicular notch, which would likely be difficult to resolve from the observations. This is because CRS's omnidirectional and directional rates are a mixture of temporal (48-s) and spatial averaging. In addition, the infrequency of spacecraft maneuvers (~6 times/year) adds to the statistical limitations of the directional data. Nonetheless, we argue that a missing notch in the pitch angle distribution is clearly the best first-order approximation to any more complicated distribution.

To obtain the results in the following section, we consider two approaches for setting limits on the notch's characteristics: 1) an empty notch (Model #1: variable width, 100% depth) and 2) a partially-filled notch (Model #2: variable width and depth). Appendices A & B describe the empty and partially filled notch models, respectively. In both cases, we define the model pitch angle distributions and determine the detailed instrument response to each type of distribution.

# 4. Results

## 4.1 Model #1 Results and Comparison with Observations

We show the results from Model #1's best fits to HET 1's roll maneuver observations in Figure 4 (see also Table 3 of Appendix A). The data was taken during 25 maneuvers that occurred from late 2012 (shortly after the heliopause crossing) to the end of 2016, when the number of rolls per maneuver was reduced. The effective widths range from 0° to ~4° (Figure 4b). Overall, this model agrees well with respective 70°-offset (Figure 4c) and omnidirectional (Figure 4d) observations for HET 1.

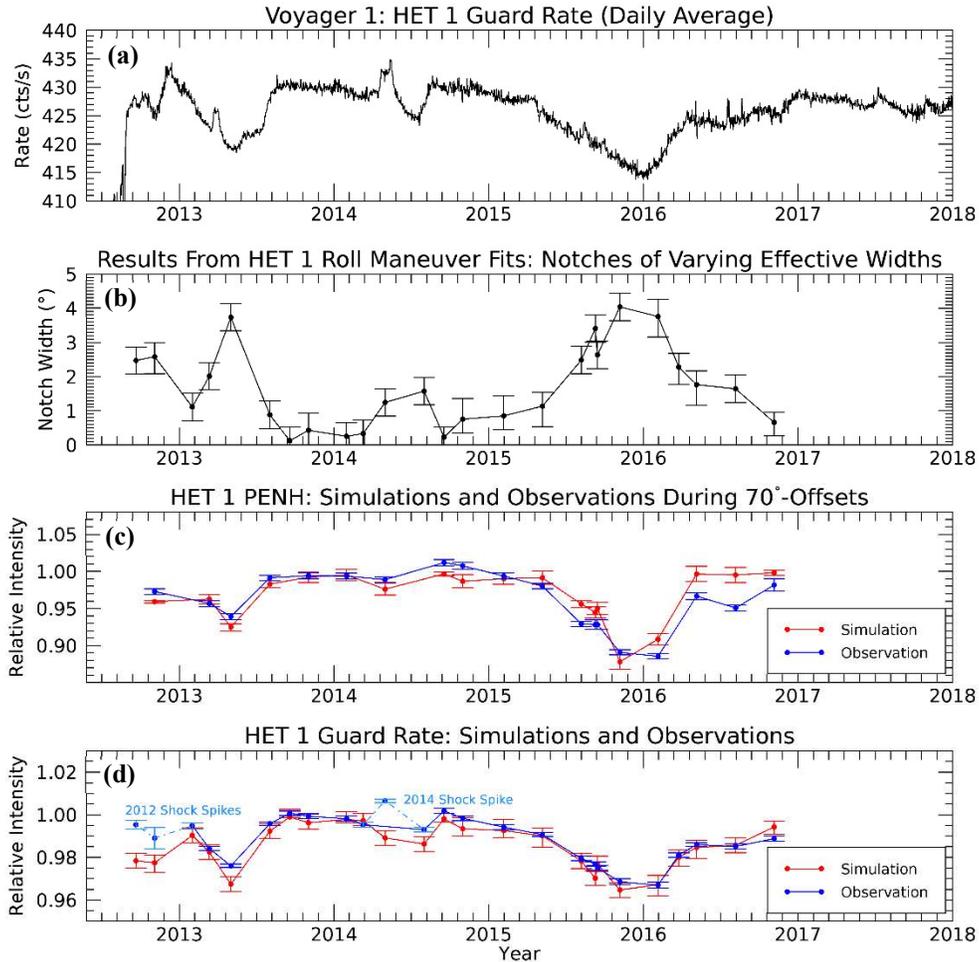

**Figure 4.** Results from Model #1 simulations and comparison with observations: HET 1.
(a) HET 1's omnidirectional guard rate (≳ 20 MeV; proton-dominated) shows the time-varying GCR intensity reductions caused by the pitch-angle anisotropy.
(b) Effective notch widths from fits to HET 1's bi-directional PENH rate (≳ 70 MeV; proton-dominated) during 25 roll maneuvers from late 2012 through 2016. We use these widths to generate the results shown in panels (c) and (d).
(c) 70°-offset simulations and observations near the 25 roll intervals for HET 1's bi-directional PENH rate (≳ 70 MeV; proton-dominated). Observed intensities are normalized to temporally-adjacent non-offset rates, while simulated intensities are normalized to modeled response function values without a notch.
(d) Omnidirectional simulations and observations during the 25 roll intervals for HET 1's guard rate (≳ 20 MeV; proton-dominated). Observed intensities are normalized to the average values during the 2013.6 to 2014.1 period when count rates are relatively uniform and isotropic, while simulated intensities are normalized to modeled response function values without a notch.

Similar results derived from the best fits to HET 2's roll-maneuvers are shown in Figure 5 and listed in Table 4 of Appendix A. HET 2's widths also vary from 0° to ~4° (Figure 5b) and results from Model #1's simulated omnidirectional notch response function agree with observations (Figure 5d). However, the 70°-offset results are not consistent (Figure 5c) with the results of this model. According to our simulation, HET 2 should not observe an intensity change, yet it observes small, but still statistically-significant anisotropic decreases.

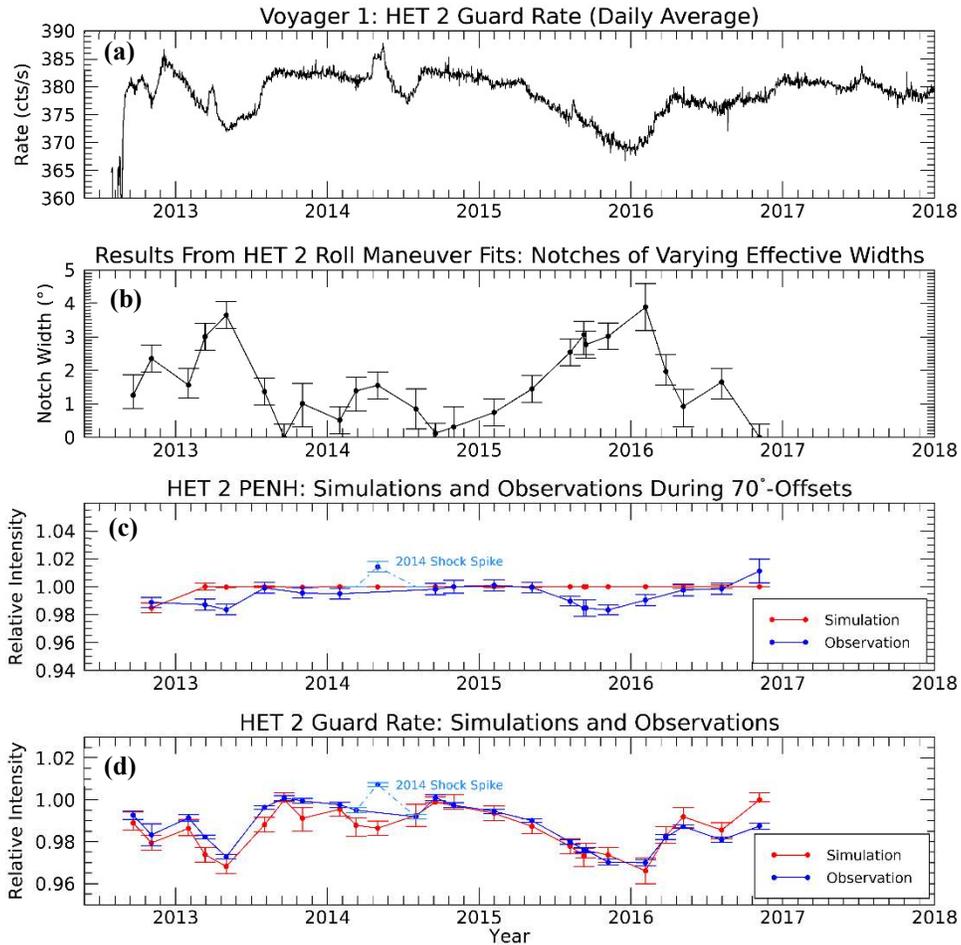

**Figure 5.** Results from Model #1 simulations and comparison with observations for HET 2.
(a) HET 2's omnidirectional guard rate (≳ 20 MeV; proton-dominated).
(b) Effective notch widths from fits to HET 2's bi-directional PENH rate (≳ 70 MeV; proton-dominated) during 25 roll maneuvers from late 2012 through 2016.
(c) 70°-offset simulations and observations near the 25 roll intervals for HET 2's bi-directional PENH rate (≳ 70 MeV; proton-dominated).
(d) Omnidirectional simulations and observations during the 25 roll intervals for HET 2's guard rate (≳ 20 MeV; proton-dominated).

Resolving this 70°-offset discrepancy requires a shift in boresight pitch angle of ~8°, which theoretically might be explained by considering uncertainties in the telescope's assumed pointing direction and the measured magnetic field direction. However, the adjustment needed is too large to be attributed to uncertainty in telescope orientation and is also beyond the range of the magnetometer's uncertainties. An added complication is that changing the magnetic field direction also affects the results for HET 1.

Instead, the most likely way to resolve HET 2's inconsistencies is by allowing for a wider notch. However, we cannot achieve this without generating inconsistencies in HET 1 if we maintain Model #1's 100% depth assumption. For example, HET 2's observations (1.7% ± 0.4%) during the 2015-296 offset interval (2015-310 roll interval) can be reproduced by an effective notch width of 19.1° ± 0.8°. Yet, this same width applied to HET 1 yields a simulated 56.2% ± 2.2% relative intensity reduction compared its 11.0% ± 0.3% observation. Thus, we resolve these issues by introducing a variable depth parameter through Model #2's broader, partially-filled notch.

## 4.2 Model #2 Results and Comparison with Observations

We report notch widths and depths from Model #2's best fits to HET 1 & HET 2 roll maneuver observations in Figure 6, focusing on the 6 intervals where the anisotropy is most prominent (see Figure 5b and Appendix B, Tables 7 & 8)

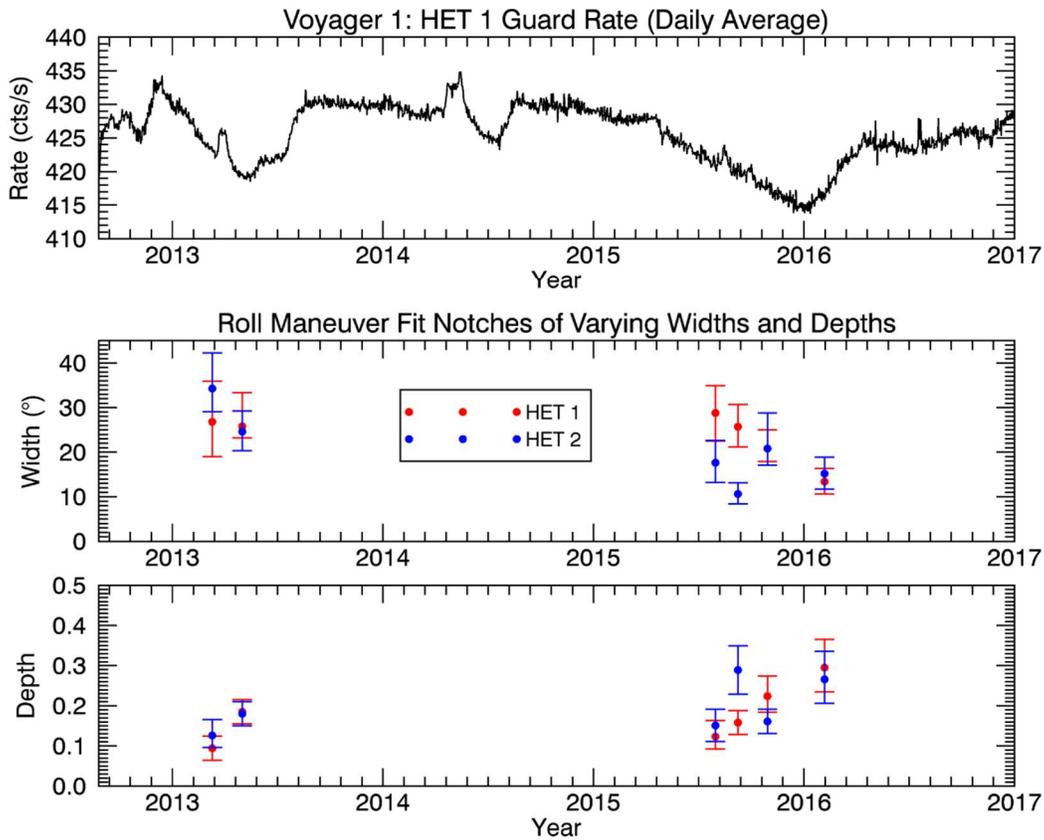

**Figure 6.** Widths and depths predicted from HET 1 (red) and HET 2 (blue) roll maneuver fits for notches of varying widths and depths. Error bars denote 1-σ uncertainties. We note that the depths signify the relative intensity deviation from isotropy; e.g. a depth of 0.1 means that the intensity within the notch is reduced by 10% compared to the isotropic baseline (see Appendix B for more details).

HET 1 & 2 results agree for 5 out of the 6 intervals. HET 2's 2015-252 offset (~2015.69) had a poor fit with a P-value of 0.50% (Table 8, Appendix B). Perhaps a small shock enhancement that occurred near 2015-224 or plasma oscillations that began on 2015-247 contributed to this outlier.

As an independent verification of the notch's parameters, we superimpose simulated HET 1 & 2 70°-offset and omnidirectional response function curves to constrain widths and depths (detailed in Appendix B). To illustrate, Figure 7 displays a superposition of HET 1 & 2 omnidirectional and 70°-offset response function curves for the 2013-120 offset. In general, HET 1 & 2 agree for shallow, broad notches (e.g. around 24° wide and 12% deep) as opposed to narrow, deep notches (e.g. towards Model #1's 100% depth limit).

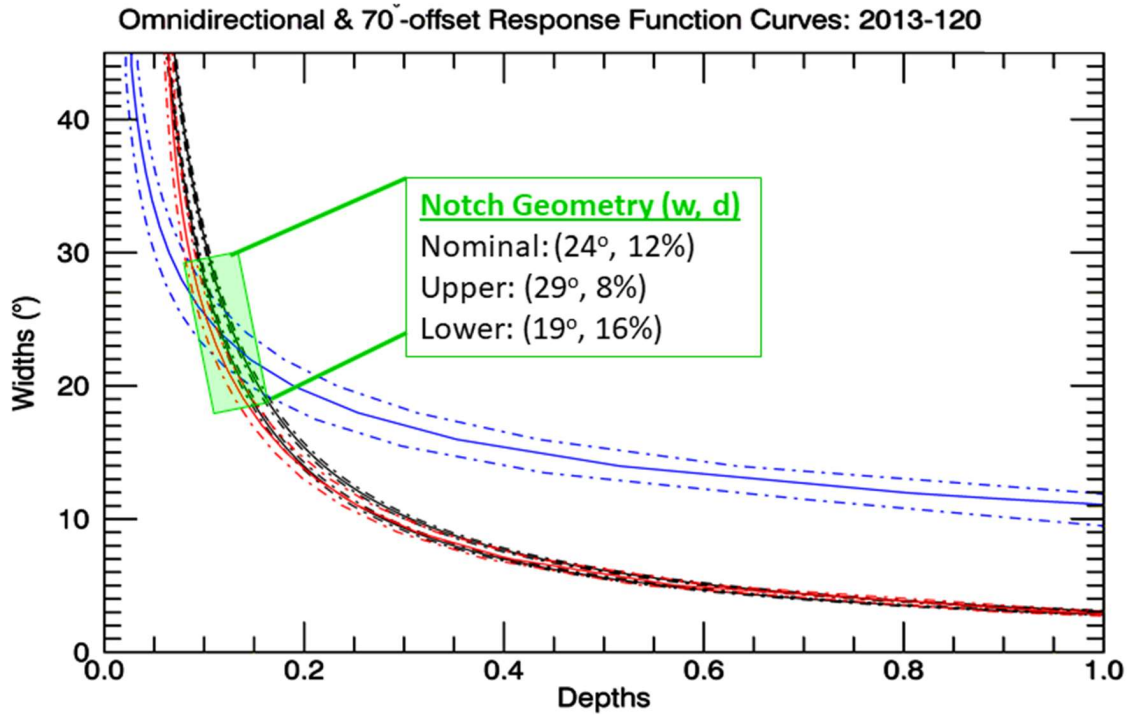

**Figure 7.** Superposition of HET 1 & HET 2 omnidirectional (black dashed) and 70°-offset response function curves (HET 1 in red, HET 2 in blue) for the 2013-120 offset interval. Note that agreement is achieved by a broad, shallow notch as opposed to a narrow, deep notch (e.g. at the limit of Model #1).

A complication of this approach is that HET 1's omnidirectional and 70°-offset curves do not always intersect due to the combination of uncertainties in the reported B-field and the particular sensitivity of HET 1's response function to pitch angle. To resolve the discrepancy for the affected offset intervals, we select field values that fall within the magnetometer's uncertainties by minimizing each component's deviation from reported observations (see Appendix B, Section B.2 for further details). Figure 8 shows the resulting notch geometries.

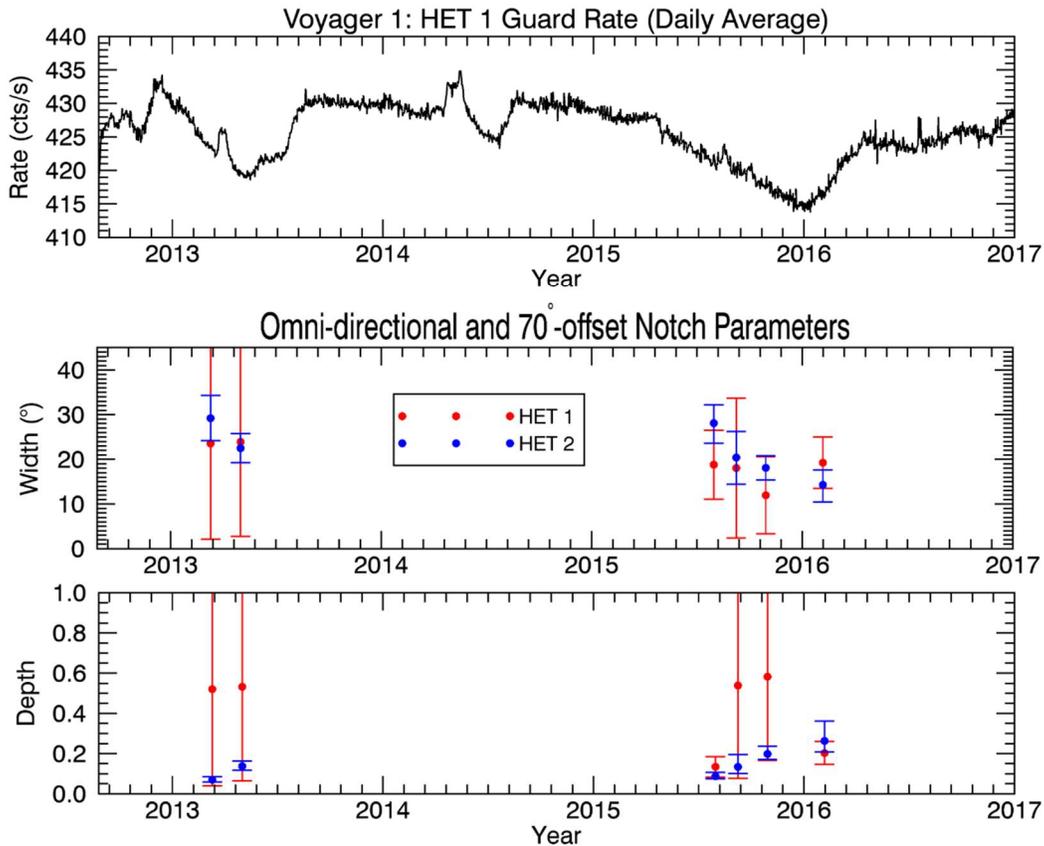

**Figure 8.** Widths and depths predicted from the intersection of omnidirectional and 70°-offset simulations for HET 1 (red) and HET 2 (blue) incorporating the alternative fields listed in Table 6 of Appendix B.

For four intervals, HET 1's 70°-offset curves matched the omnidirectional curves over a broad range of widths and depths, and therefore could not effectively constrain the notch's geometry. However, for two intervals (2015-208 and 2016-31), they were sufficiently different to allow HET 1 to confirm the broad, shallow notch seen by HET 2. Regarding uncertainties, we note that preserving the observed notch areas ($\delta_{omni}$ and $\delta_{70°}$) causes width and depth to vary inversely proportionally to each another. In other words, the wider the notch, the shallower the depth.

In Figure 9 we compare HET 1's roll-maneuver notch parameters to the results from HET 2's 70°-offset and omnidirectional response function curves. The independently-acquired results from these two approaches show agreement, and HET 1 & 2's widths and depths are consistent with one another favoring a broad, shallow notch that is, on average 22° wide and 15% deep.

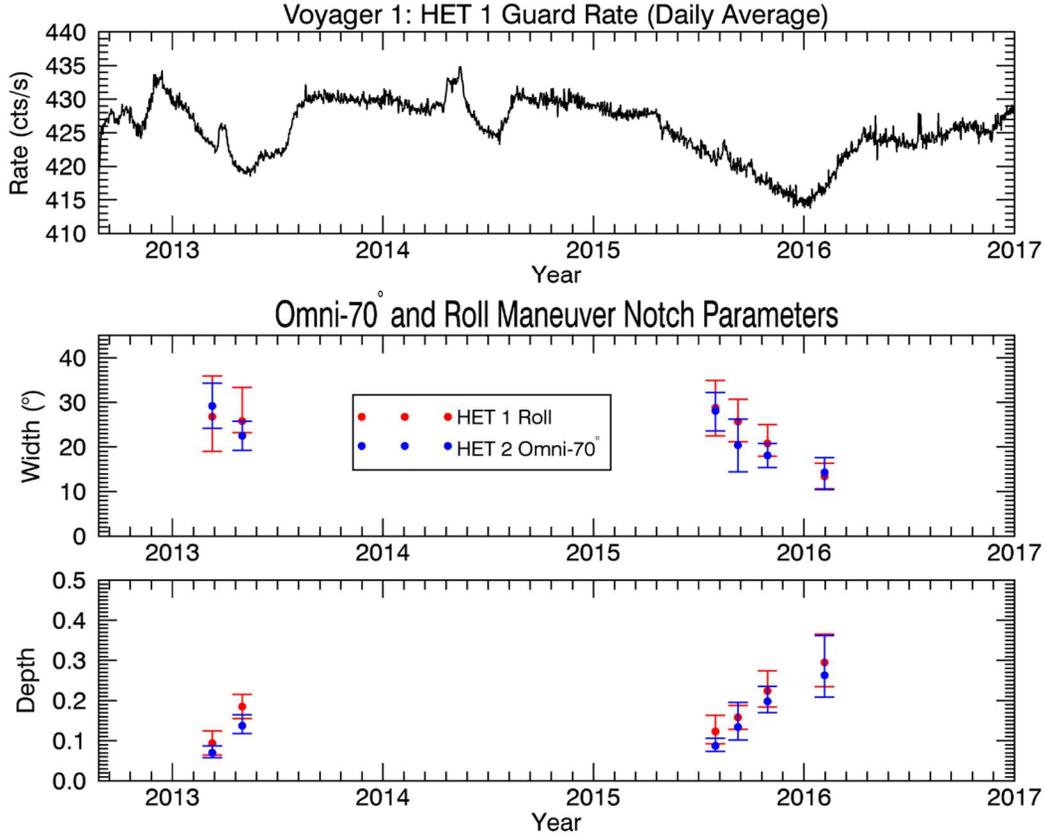

**Figure 9.** Comparison of HET 1 roll maneuver widths and depths (red) to independently-acquired HET 2 omni-70° results (blue).

## 5. Discussion
### 5.1. Anisotropy Formation via Magnetic Trapping
To understand the underlying physics of the pitch angle anisotropy, we must consider both its spatial formation and temporal evolution. The cosmic rays that are considered throughout this work are primarily on the order of several-hundred MeV (Cummings et al. 2016) and are observed in the LISM where their scattering mean free paths are very large (Ptuskin 2001). Additionally, magnetic fluctuations beyond the heliopause are very small (Burlaga et al. 2018). Therefore, it is reasonable to assume that particle energies remain constant as they follow slowly-varying magnetic field lines. As such, we can describe the anisotropy's spatial formation through magnetic trapping, which arises from the conservation of the first adiabatic invariant:

$$\frac{\sin^2 \alpha(x)}{|B|_x} = \frac{\sin^2 \alpha_o}{|B|_o} = const. \quad (1)$$

for a distance $x$ along the field line and values $\alpha_o$ and $|B|_o$ at the point of observation. As particles with pitch angles $\alpha(x)$ encounter stronger fields, their pitch angles increase until they reach a mirror point $(x_m)$ – where $\alpha_m = 90°$ and $|B|_m = \frac{|B|_o}{\sin^2 \alpha_o}$ – after which the parallel component of the Lorentz force causes particles to reverse direction and move back towards regions of lower $|B|$. Therefore, if a weak magnetic field ($|B|_w$) is bounded by a strong field in either direction along the field line ($|B|_s$ – although the two directions could differ in strength), the largest pitch angles to arrive in the weaker field region ($\alpha_w$) will be determined by:

$$\alpha_w = \sin^{-1}\left(\pm\sqrt{\frac{|B|_w}{|B|_s}}\right) \quad (2)$$

for particle velocities parallel ($+$) and anti-parallel ($-$) to the field. Hence, $\alpha_w < 90°$ because $\sqrt{\frac{|B|_w}{|B|_s}} < 1$, resulting in a pitch angle gap in the weak field region near 90° with a total width of:

$$w = 2(90° - \alpha_w) \quad (3).$$

Since trapping only occurs within a region bounded by stronger fields, what, then produces this requisite field geometry?

### 5.2. Trapping Mechanisms

Figure 10 illustrates three possible trapping scenarios. The boundaries of a quasi-perpendicular shock could, in theory, supply the necessary geometry for trapping (Figure 10a). According to Kóta & Jokipii (2017), particles trapped downstream (in the shocked plasma) undergo cooling as they interact with adiabatically-expanding magnetic fields (see Figure 1 of Kóta & Jokipii 2017). Those with pitch angles nearest to 90° stay trapped for the longest amount of time, experience the greatest energy loss, and therefore contribute the most to reductions in GCR flux (see Figures 3 & 5 of Kóta & Jokipii 2017).

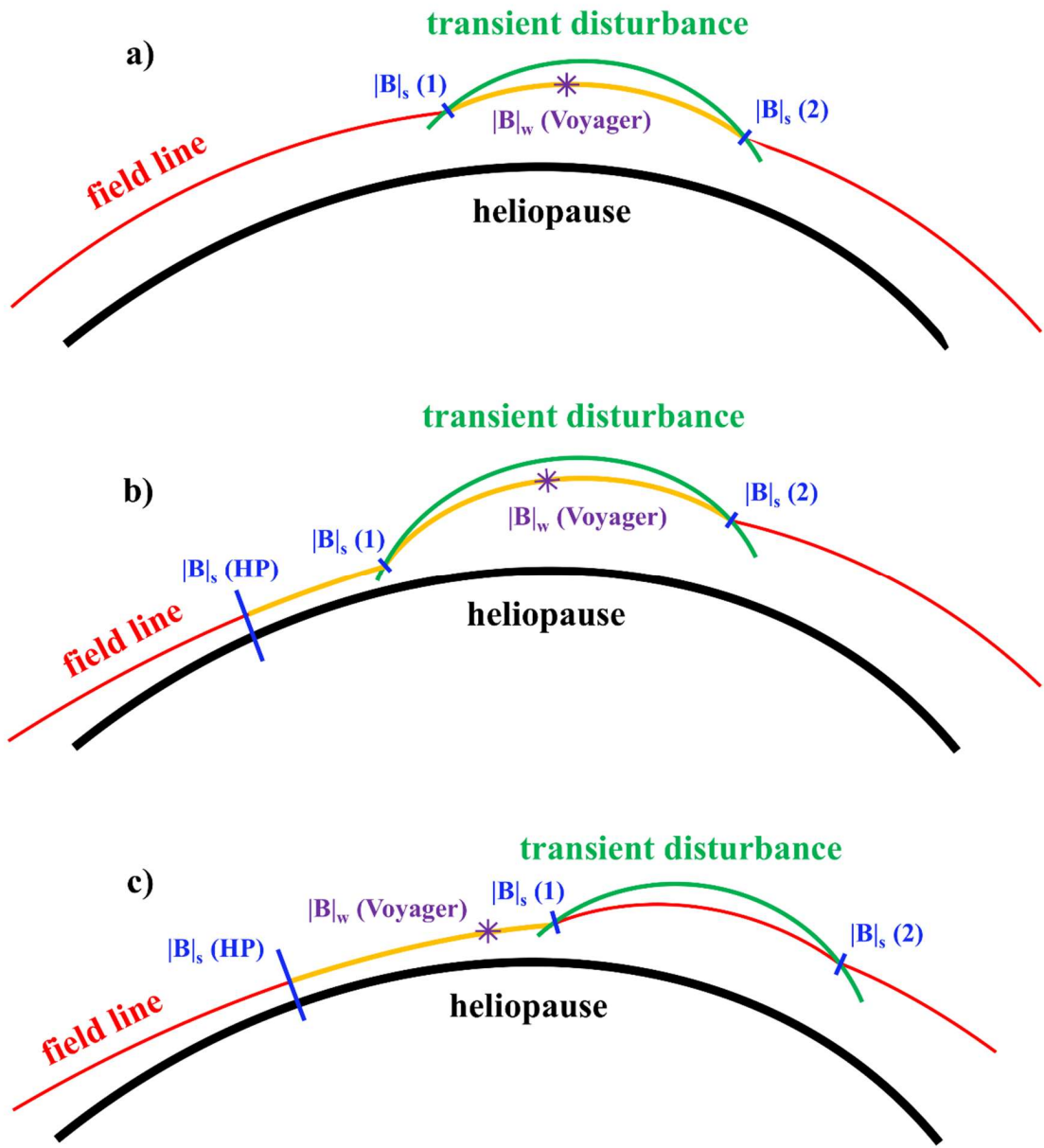

**Figure 10.** Three possible scenarios that produce the necessary geometry for particle trapping; each consists of strong fields ($|B|_s$, blue) bounding a weak field at *Voyager* ($|B|_w$, purple).
a) Condition suggested by Kóta & Jokipii (2017) in which particles are trapped completely within a transient disturbance (green). The trapping region (between strong fields $|B|_s(1)$ & $|B|_s(2)$) is shown in orange.
b) Similar to a), but also including the field compression towards the heliopause, $|B|_s$(HP). This is one possible illustration; the key points are that 1) trapping occurs between maxima along the field line and 2) IBEX observations and models strongly support the existence of the more steady-state field compression toward the heliopause. Depending upon the relative strengths of $|B|_s(1)$, $|B|_s(2)$, and $|B|_s$(HP) the particles could remain trapped inside the disturbance or the trapping region could extend between the heliopause and the disturbance.
c) An alternative condition in which the disturbance does not cross Voyager, but particles are trapped between the heliopause compression, $|B|_s$(HP), and the disturbance (the stronger of $|B|_s(1)$ or $|B|_s(2)$).

Many aspects of this model are compelling. For example, Kóta & Jokipii (2017) predict decreasing trends in the GCR flux for a μ = 0.00 to 0.25 pitch angle segment and not in other segments (μ > 0.25) (see their Figure 3 & 5). This translates to a maximum notch width of: $w = 2 \times (90° - \cos^{-1}(0.25)) = 29°$. Our analysis agrees. Model #2 prescribed a ~29°-wide notch for 2013-67 (Table 10 in Appendix B); this is the interval nearest to the shock that Kóta and Jokipii used to inform their model, which was observed by *Voyager* on ~2012-335 (see Burlaga & Ness 2016).

The global structure of the heliopause likely also influences the trapping geometry, as shown in Figures 10b & 10c. We argue here that the draping and compression of the interstellar magnetic field around the heliopause, as demonstrated by IBEX, naturally creates a permanently compressed field region that could also serve as a mirror point for the trapped particles. The steady-state field compression likely occurs where the field is most tangent to the heliopause: at the IBEX ribbon. Indeed, according to Zirnstein et al. (2016)'s model, the field near the heliopause is strongest in the direction of the ribbon and is ~20% stronger (private communication with E. Zirnstein) than the 0.48 nT average field seen by *Voyager* (Burlaga & Ness 2016).

As Figure 10b illustrates, the situation described by Kóta & Jokipii (2017) and the existence of the naturally steady-state compression are not necessarily exclusive, depending on the strong-field boundary conditions. For example, if $|B|_s(HP) < |B|_s(1)$ & $|B|_s(2)$, particles will remain trapped completely within the disturbance (as in Figure 10a). Alternately, if $|B|_s(1) < |B|_s(HP)$, the trapping region could extend towards the heliopause compression region . We also note that the boundary conditions could change over time; even if the trapped particles are initially contained within the disturbance, they could eventually mirror at $|B|_s(HP)$ as the transient weakens over time or distance.

Figure 10c shows another possible scenario where *Voyager* remains in the un-shocked plasma without locally encountering the disturbance itself. If $|B|_s(1) < |B|_s(2)$, *Voyager* could still observe the cooled particles as they escape from the disturbance. Alternatively, anisotropies could also form if the locally-trapped particles (e.g. those near 90°-pitch angle) are affected by some (perhaps unrelated) microscopic or macroscopic energy loss mechanisms such as expanding fields produced by an inward motion of the heliopause (Washimi et al. 2011, 2017).

### 5.3. Physical Interpretation of Pitch Angle Notches

We now consider how particle trapping relates to the notch's width and depth. Suppose *Voyager* is on a field line that is filled with an isotropic distribution of particles and has an enhanced strength near the heliopause compression but no other local maxima. So long as these initial conditions are satisfied, *Voyager* will continue to observe particles at all pitch angles with constant intensity. When a compression passes by, it generates additional local maxima. This change of boundary conditions isolates part of the pitch angle distribution (those near 90°) so that the affected particles are no longer replenished by the surrounding vast cosmic ray reservoir. The notch's width is a measure this isolation. However, trapping alone is not sufficient to produce a notch; either particles must leak out of the trap more quickly than it is re-filled or they must experience some sort of energy loss that translates to a reduction in intensity. In other words, the notch represents a combination of both width and depth, trapping and intensity change.

The notch's width reflects the extent of the affected particle pitch angle distribution and is a measure of the ratio of the strong and weak fields. Combining Equations 2 & 3 leads to:

$$\Delta |B|_n = (|B|_s - |B|_w)/|B|_s = \cos^2(\alpha) = \cos^2(90° - w/2) \quad (4)$$

where $\Delta |B|_n$ is the change in the ambient field that is required to produce a notch of width $w$. A notch that is too narrow could potentially be erased by turbulent fluctuations in the steady-state magnetic field. In the LISM, *Voyager* observes fluctuations of ~2% over several-week timescales (Burlaga et al. 2015; Burlaga et al. 2018); given that the GCR anisotropic decreases endure for many months at a time, this serves as a lower limit to $|B|_n$. Additionally, a notch should not require changes in the field that are larger than we observe. For example, in 2014, *Voyager* encountered a traveling shock that produced a ~12% enhancement of the field (Burlaga & Ness 2016); this provides an upper limit to $\Delta |B|_n$. Thus, we expect the notch widths from our simulations to reflect field changes of ~2% $\lesssim \Delta |B|_n \lesssim$ ~12%. Averaging Model #2's six intervals yields a notch that is 22° wide and 15% deep (Section 4.2), so $\Delta |B|_n = 4\%$. Since 2% < 4% < 12%, our average broad, shallow notch implies changes in $|B|$ that are reasonably consistent with the observations.

The notch's depth reveals how magnetic trapping leads to changes in the GCR intensity. Two likely contributors include adiabatic cooling and scattering. When *Voyager* is downstream of the shocked plasma, adiabatic cooling is likely the dominant energy loss mechanism (Figures 10a & 10b) as suggested by Kóta & Jokipii (2017). Due to the negative spectral index of the relevant-energy GCRs in the LISM (Cummings et al. 2016), loss in particle energy translates to reduced intensity, per Liouville's theorem (see for example, Kóta & Jokipii 2017). A partially-filled notch could also be indicative of some non-adiabatic process. For example, the local turbulence – especially if it differs from its surroundings – might affect the rate at which particles escape (or enter) the trapping region. The interplay between turbulence and cooling might also provide clues as to why these notches are long-lasting, yet "mostly-filled". The relative amount that adiabatic cooling or scattering processes may contribute to (or hinder) the notch's formation merits further study. The depth likely grows as a function of the time that the particles spend in the trap. Again, we emphasize that our best-fit notches are mostly filled – on average only 15% deep – and the changes we measure in the omnidirectional intensity distribution are small – less than 4% – so only small amounts of cooling or scattering are necessary to produce the effects that we observe.

Figure 11 shows a schematic B-field diagram of a trapping scenario where the fields of the disturbance are weaker than the compression towards the heliopause. Again, *Voyager* must be in the weak field region in order to observe the anisotropy and the region must be bounded by stronger fields along those field lines. The compressed field at the heliopause likely differs in magnitude from the enhanced field at the disturbance, so if both of these contribute to the trapping, it is the weaker of these strong fields that ultimately determines the notch's width.

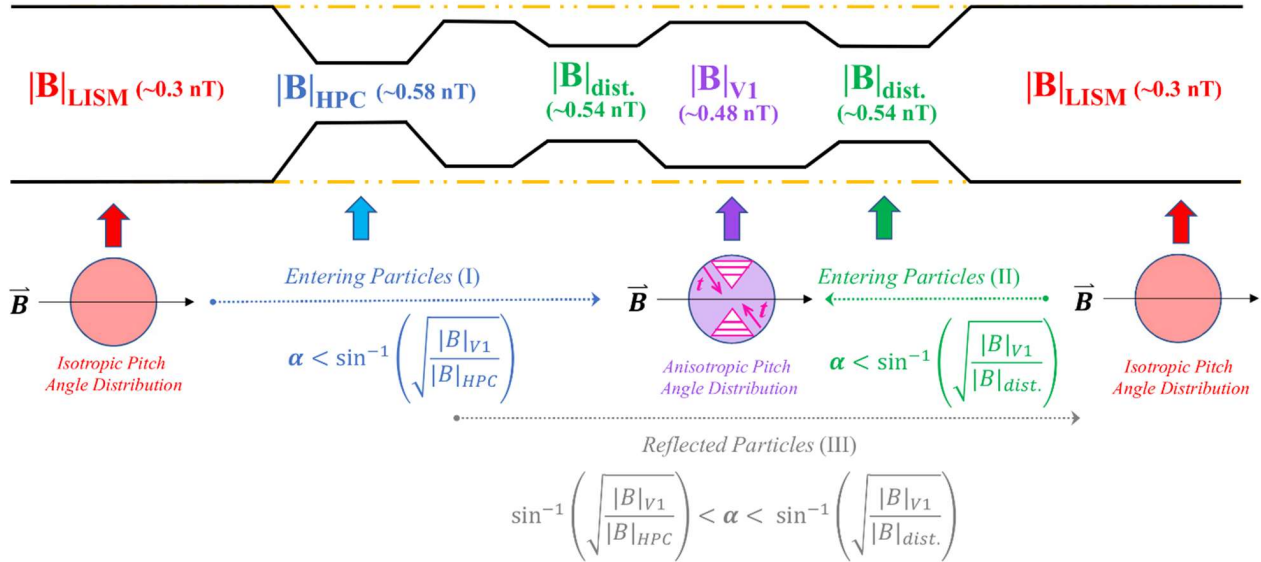

**Figure 11.** Illustration of a flux tube (black lines) showing a magnetic field and particle configuration for which the field towards the heliopause compression is stronger than at the disturbance ($|B|_{HPC} > B|_{dist}$). Particles originate from an isotropic distribution at either end ($|B|_{LISM}$ (red)). The notch forms as they get trapped by the enhanced fields at (I) the steady-state compression near the heliopause ($|B|_{HPC}$, blue) and (II) the temporary compression of the disturbance ($|B|_{dist.}$, green). The weaker of the strong fields sets the limit to the notch's width ($|B|_{dist.} < |B|_{HPC}$ in this example). This is because some of the particles that pass through the weaker compression are later reflected when they encounter the yet stronger field (III, grey). Intensities change as particles lose energy in the adiabatically-expanding fields, or possibly if they experience preferential scattering due to turbulence. The notch's depth is a function of the amount of time that the particles are trapped (center circle, pink). The $|B|_{LISM}$ field strength is that of the unperturbed LISM at >1,000 au (Zirnstein et al. 2016), $|B|_{HPC}$ is ~20% of the field at *Voyager* (private communication with E. Zirnstein), $|B|_{V1}$ reflects the average value seen at Voyager, and $|B|_{dist.}$ reflects the magnitude of the compression caused by the transient event that crossed *Voyager* on ~2014-237 (Burlaga & Ness 2016).

Figure 12 shows the sequence of LISM shock transient events observed by *Voyager* on multiple instruments. GCR's periodically undergo roughly month-long intensity enhancements (e.g. Figure 12a, near ~2013.2 and ~2014.3) which are reminiscent of the shorter-lived particle spikes produced at the foreshock of interplanetary shocks, as modeled by Jokipii and Kóta (2014) and noted by Gurnett et al. (2015). Due the particles' high energies, these are the first indications of the transients to arrive at Voyager. Next, *Voyager 1*'s Plasma Wave Subsystem observes emissions from electron plasma-beam instabilities, which also occur upstream of the disturbances (Gurnett et al. 2015; horizontal bars in Figure 12b). Finally, several weak, smooth, thick disturbances cross *Voyager* and are measured by the magnetometer (Burlaga & Ness 2016; vertical lines in Figures 12a & 12b).

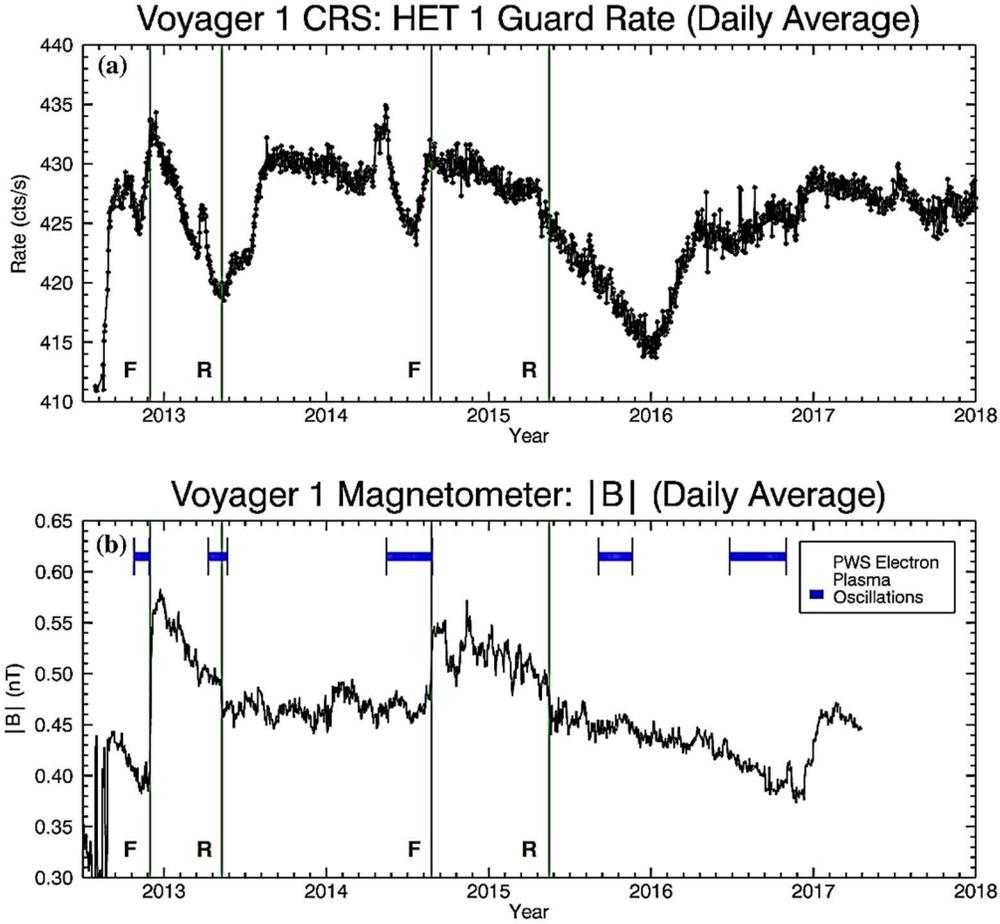

**Figure 12.** Comparison of shock transient and GCR anisotropy events observed by *Voyager 1* in the LISM.
(a) HET 1's omnidirectional guard rate (≳ 20 MeV; proton-dominated) with vertical lines indicating forward and reverse disturbances observed by *Voyager 1*'s magnetometer (see Burlaga & Ness 2016).
(b) Magnetometer B-field strength vs. time with vertical lines indicating the disturbances as in panel (a) (from the magnetometer's publicly-available data: https://omniweb.gsfc.nasa.gov/coho/form/voyager1.html). The horizontal blue bars indicate the timing of emissions from electron plasma oscillations recorded by the Plasma Wave Subsystem (PWS)(see Gurnett et al. 2013 and Gurnett et al. 2015 for further details). The time periods shown reflect those of the 3.11 kHz channel electric field measurements (from publicly-available PWS data: http://www-pw.physics.uiowa.edu/voyager/data/).

We compare the trends of our model to the observations (Figure 12) to evaluate our physical interpretation of the pitch angle notches. Table 2 contains a listing of Model #2's results, along with an estimate of the magnitude of the strong field ($|B|_s$) that is involved in producing the various notches. These estimates are derived from Equation 4 by substituting Voyager's local magnetic field measurements for the weak field ($|B|_w = |B|_{obs}$).

| Interval | width | depth | $\delta_{omni}$ | $\Delta|B|_n$ | $|B|_{obs}$ | $|B|_s$ |
|---|---|---|---|---|---|---|
| **2013-67** | 29° ± 5° | 7% ± 1% | 1.8% ± 0.05% | 6% ± 1.3% | 0.52 ± 0.04 nT | 0.55 ± 0.12 nT |
| **2013-120** | 23° ± 3° | 14% ± 2% | 2.7% ± 0.05% | 4% ± 0.6% | 0.49 ± 0.04 nT | 0.51 ± 0.10 nT |
| **2015-208** | 28° ± 4° | 9% ± 2% | 2.1% ± 0.05% | 6% ± 1.1% | 0.46 ± 0.04 nT | 0.49 ± 0.10 nT |
| **2015-250** | 20° ± 6° | 13% ± 5% | 2.4% ± 0.07% | 3% ± 1.1% | 0.45 ± 0.04 nT | 0.46 ± 0.17 nT |
| **2015-296** | 18° ± 3° | 20% ± 3% | 3.1% ± 0.05% | 2% ± 0.4% | 0.45 ± 0.04 nT | 0.47 ± 0.09 nT |
| **2016-31** | 14° ± 4° | 26% ± 8% | 3.3% ± 0.06% | 2% ± 0.4% | 0.43 ± 0.04 nT | 0.44 ± 0.13 nT |

**Table 2.** B-field variations ($\Delta|B|_n$) required to produce a notch of a given geometry, informed by Model #2's parameters. $|B|_{obs}$ is the strength of the local field measured by Voyager. $|B|_s$ represents the magnitude of the strongest field in the trapping configuration. The latter is determined by substituting the observed field ($|B|_{obs} = |B|_w$) into Equation 4 (see also Appendix B, Tables 5, 6 & 10).

The largest predicted $|B|_s$ in Table 2 occurs during the 2013-67 epoch, and its ~0.55 nT value is very close to the reported maximum field strength (~0.56 nT) of the magnetometer's 2012-335 shock transient event (~2012.9 in Figure 12b; see also Figure 3 in Burlaga & Ness 2016). During the 2015-208 epoch, $|B|_s \sim 0.49$ nT, which agrees with the 0.494 nT field observed at the time of the preceding 2015-137 reverse shock (~2015.4 in Figure 12b; see also Figure 5 in Burlaga & Ness 2016). Since our calculated strong field seems consistent with the strength of passing disturbances, the steady-state field towards the ribbon is likely stronger than the temporary fields for these periods.

Table 2's trends in widths and depths (per episode) make sense in relation to the disturbances' temporal behavior. We expect the magnetic field (hence the widths) to weaken towards the anisotropy minima (intervals 2013-120 and 2016-21 in Table 2; ~2013.35 & ~2016.0 in Figure 12a) since the field at the transients likely weakens as the shock moves further out. Additionally, we expect the depth – and therefore the magnitude of the anisotropy – to grow as a function of longer cooling and scattering times, so long as the particles remain trapped.

There are local indications that adiabatic cooling plays an important role in the anisotropy's growth. For example, in 2013, the field weakens on the same time scales that the anisotropy develops (see ~2012.9 to ~2013.35 in Figure 12). The 2015 episode shows similar behavior, although not identical, since its development occurs in a two-step process (from ~2014.65 to ~2015.35 and ~2015.35 to ~2016.0). The 2014 episode, however, is an exception; it occurs during a time when the local fields appear neither expanded nor compressed and is not preceded by an obvious disturbance. Perhaps the field geometry is similar to Figure 10c and *Voyager* senses cooled particles escaping from the shock or some different energy-loss mechanism is affecting the locally-trapped particles.

The onset of each recovery seems to occur when the trap has either dissipated, or *Voyager* has moved beyond the trapping region. For example, the 2013 episode recovers while the local field no longer weakens but appears to fluctuate about ~0.46 nT (~2013.35 to ~2013.6 in Figure 12). For the 2015 episode, Table 2 shows that the strong field weakens to roughly the observed value by 2016-31 ($|B|_{obs} \sim |B|_s$), near the GCR intensity minimum (~2016.0 in Figure 12a). However, beyond the minimum (Figure 12), the local magnetic field continued to decrease below average through the end of 2016. Perhaps during this recovery, Voyager's field lines were still connected to the shock, but no longer connected to the strong field towards the ribbon, or maybe some sort of change occurred near the heliopause.

### 5.4 Summary and Conclusion
We have provided evidence that the GCR pitch angle anisotropy observed by *Voyager 1* in the LISM is characterized by a broad, shallow, mostly-filled notch caused by particles that are missing near 90° in an otherwise uniform pitch angle distribution. We suggest that the notch forms in a trapping region that, in addition to being affected by temporarily-compressed fields from traveling disturbances, could also be affected by the presence of a steady-state enhanced magnetic field near the heliopause. IBEX observations support the existence of the former, at the heliopause near the ribbon. Regarding the latter, all of Voyager's working instruments have detected signatures of several weak, solar-induced LISM transients. The notch's width correlates with the ratio of the local field at *Voyager* and the remote field of the compressions. The notch's depth relates to the anisotropy's growth as a function of time and is at least partially due to adiabatic cooling. Topics that merit further investigation include: 1) investigating the role that turbulence might also play in the anisotropy's development, 2) examining the factors that contribute the anisotropy's magnitude and recovery, 3) understanding how the anisotropy behaves as a function of particle species and energy, and 4) exploring how the steady-state compression towards the ribbon might additionally affect the trapped particle population. We look forward to continuing our evaluations of Voyager's in-situ observations together with IBEX's global measurements, as this will potentially lead to greater insight about the heliosphere and its interaction with the interstellar medium.

## Acknowledgements

We thank E. Roelof for first suggesting particle trapping as a means for generating the anisotropy, R. Decker and the LECP team for providing background-corrected data, and E. Zirnstein for providing model-based information about the field enhancement towards the heliopause. This work was supported by NASA Grant NNN12AA01C. JSR and DM were also supported by the Interstellar Boundary Explorer (IBEX) mission, which is part of NASA's Explorer Program.

# Appendix A – Model #1: Empty Notch

Model #1 assumes negligible scattering and represents the notch as a complete dropout of particles within a range of pitch angles characterized by variable width and 100% depth (see Figure 13). These assumptions enable us to efficiently fit the simulated directional response functions to data using a single "effective width" parameter.

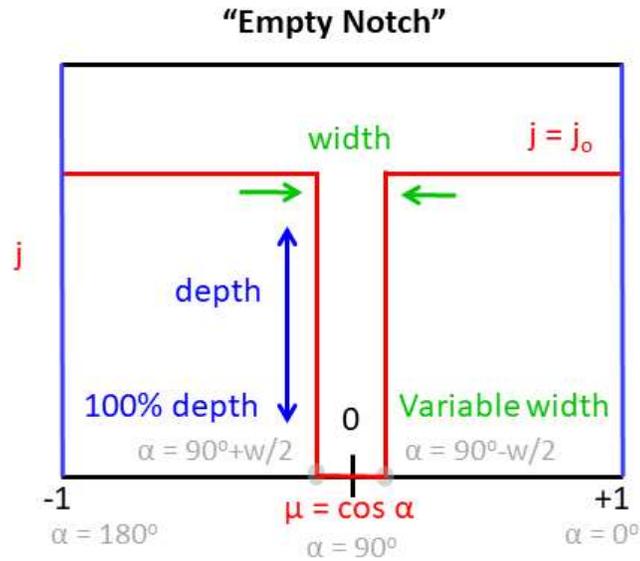

**Figure 13.** Diagram of notch Model #1.

### A.1 Omnidirectional Response Function

The omnidirectional intensity ($J$) is represented by the general expression:

$$J = 2\pi \int_{-1}^{1} j(\mu)\, d\mu \quad (A1)$$

where $j$ denotes directional particle intensity and $\mu$ is related to pitch angle, α, by: $\mu = \cos \alpha$. For an isotropic distribution, $j = j_0$ and is constant, so $J = 4\pi j_0$. For a distribution with a notch, the missing particle intensity ($J_n$) is given by the integral over the notch's effective width $w$ centered at $\alpha = 90°$:

$$J_n = 2\pi \int_{\cos(90°+w/2)}^{\cos(90°-w/2)} j\,\mu\,d\mu = 4\pi\, j_n \cos(90° - w/2). \quad (A2)$$

Assuming negligible scattering implies that $j_n = j_0$, leading to the normalized "omnidirectional notch response function":

$$\delta_{omni} = \frac{J_n}{J} = \cos(90° - w/2) = \cos\alpha. \quad (A3)$$

**A.2 Directional Response Functions**

First, we simulate a magnetometer roll maneuver by using a Monte Carlo simulation to calculate the pitch angle distribution of particles passing through HET 1 or HET 2 (detailed in the following section), accounting for each interval's observed magnetic field direction (e.g. Figures 14 & 15).

By excluding particles within an effective width centered about 90° pitch angle and normalizing to observed rates, we produce a smooth width-dependent roll maneuver notch response function (e.g. Figure 16) and calculate its $\chi^2$ with respect to the 48-s data. After repeating the process for different widths, we minimize $\chi^2$ to acquire a best-fit notch for each roll interval.

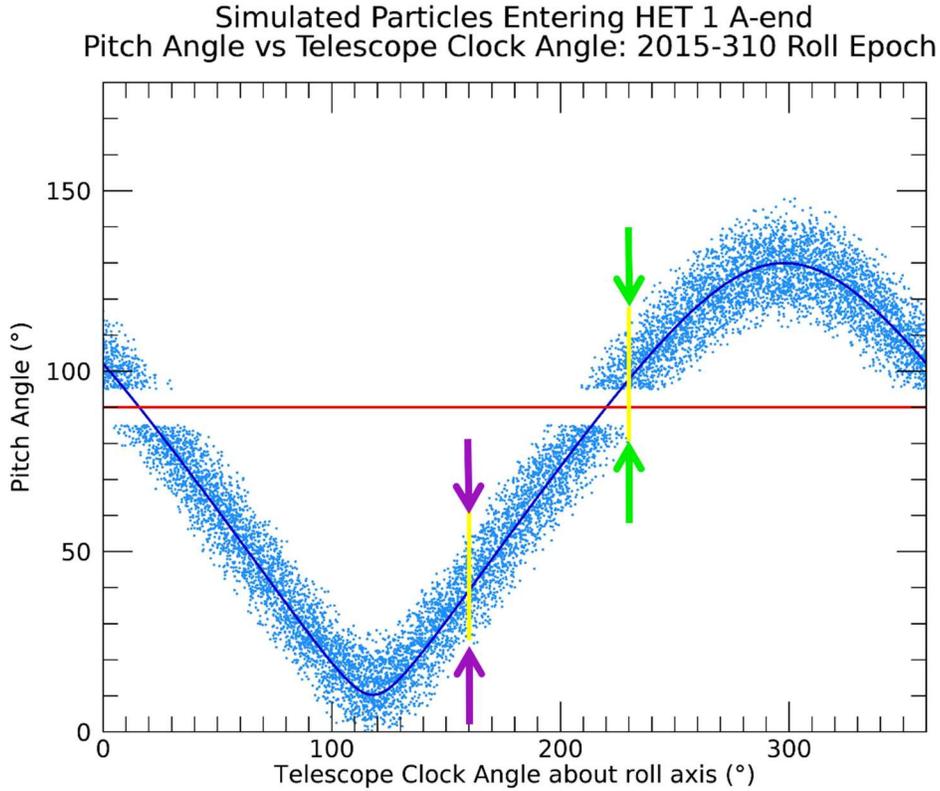

**Figure 14.** Pitch angle vs. telescope clock angle (measured from $\hat{N}$ towards $\hat{T}$) view of the 2015-310 roll maneuver Monte Carlo simulation for particles entering HET 1's A-end, shown with a 10°-wide notch. The magnetic field direction during this time was (0.156, -0.381, 0.202) nT in R, T, N (from the magnetometer's publicly-available

data: https://omniweb.gsfc.nasa.gov/coho/form/voyager1.html). Simulated particles (blue dots) fall within a ~40°-wide band, as defined by the telescope's opening angle. HET 1's nominal boresight is centered at 160.9° clock angle and 40.5° pitch angle; its normal field of view (indicated by the purple arrows) does not overlap with 90° pitch angle (red horizontal line). However, when the HET 1 boresight passes through ~17° and ~219° clock angle during the 2015-310 roll maneuver, the notch is directly centered in its field of view; therefore, a measurable count rate reduction is observed (see Figure 16a). We note that the clock-angle difference between the two pitch-angle minima is close to, but not at 180°. This is because the B-field is mostly in the $\hat{N}$ - $\hat{T}$ plane, but not quite perpendicular to the spacecraft's rotation axis ($\hat{R}$). HET 1's 70°-offset boresight is at 230.9° clock angle and 98.5° pitch angle, so its field of view also overlaps with the anisotropy during the offsets (green arrows).

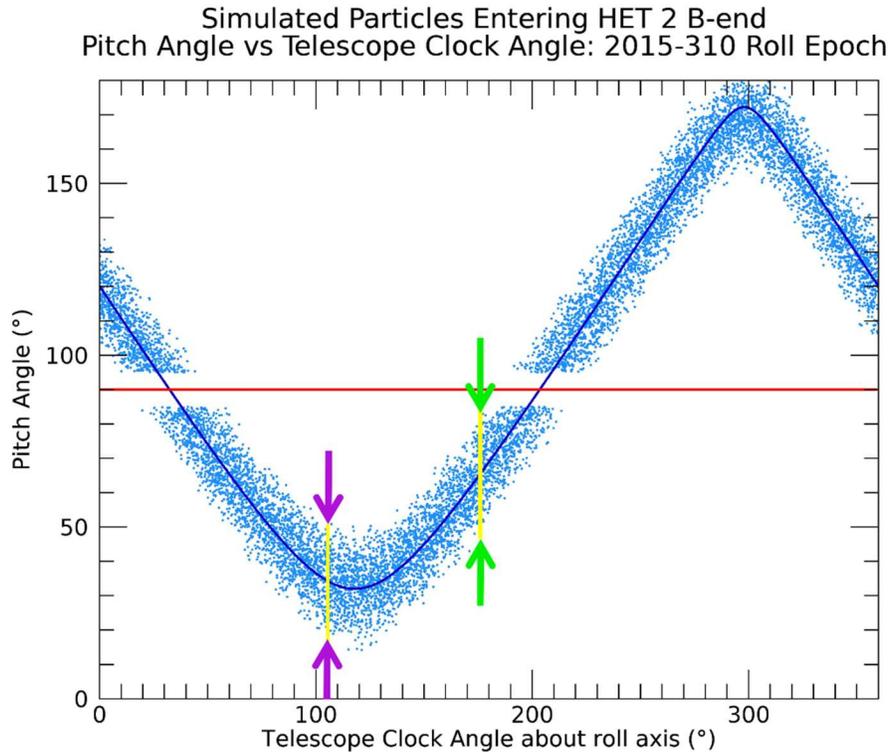

**Figure 15.** Similar to Figure 14, but for particles entering HET 2's B-end. HET 2's normal field of view (purple arrows) does not overlap with 90° pitch angle (red horizontal line) since its normal boresight is centered at 107.1° clock angle and 33.7° pitch angle. HET 2's 70°-offset boresight for 2015-310 is at 177.1° clock angle and 66.4° pitch angle, placing HET 2's field of view (green arrows) at the edge of the anisotropy; it may see an intensity decrease if the notch is wide enough.

Figure 16 shows simulated roll maneuver response function fits to observed bi-directional HET 1 PENH counts during the 2015-310 maneuver. The best fit was generated by a notch with an effective width of 4.0° ± 0.4°. Fits in both clock angle (Figure 16a) and pitch angle (Figure 16b) space yield the same effective widths (to within ±0.05° or smaller) for all intervals. Since CRS telescopes view the anisotropy as a function of clock angle during the roll maneuvers, we report clock-angle fits throughout this work.

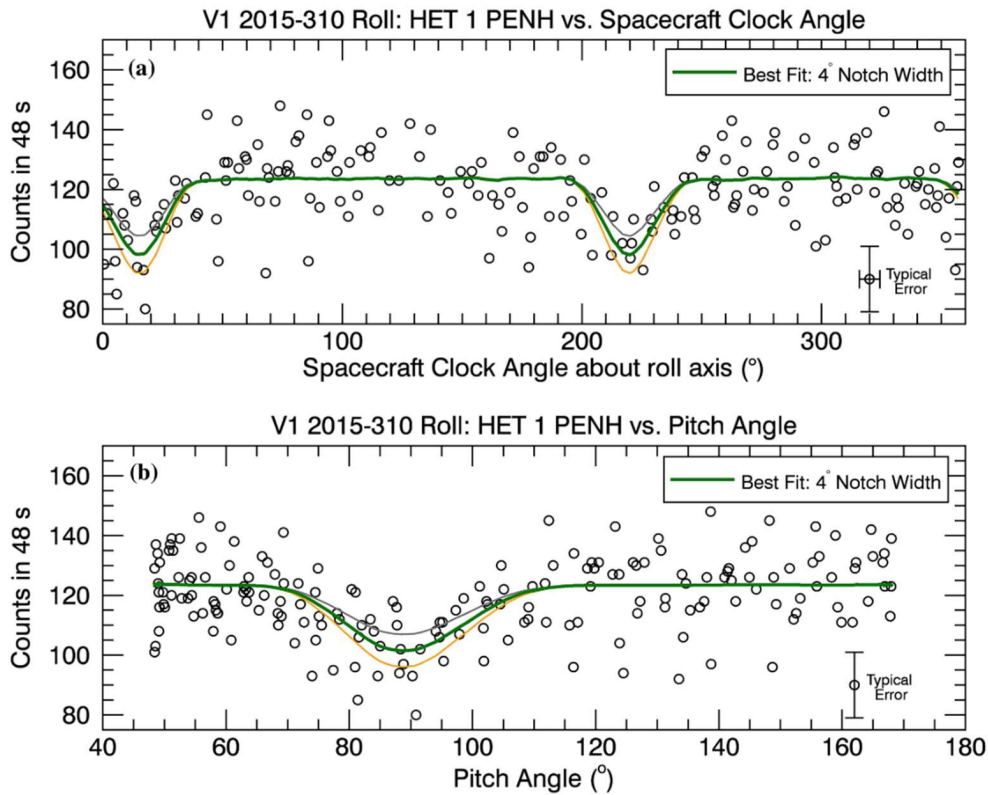

**Figure 16.** HET 1's bi-directional PENH rate (≳ 70 MeV; proton-dominated) vs. clock angle (a) and pitch angle (b) during the 2015-310 roll maneuver. The magnetic field direction during this time was (0.156, -0.381, 0.202) nT in R, T, N (from the magnetometer's publicly-available data: https://omniweb.gsfc.nasa.gov/coho/form/voyager1.html). The thick green solid line superimposed over the data represents the best-fitting notch roll response function produced by a Monte Carlo simulation with a width of 4.0° ± 0.4°. Independent fits applied in clock angle space and pitch angle space yielded the same best fit geometry. The thinner top (grey) and bottom (gold) lines represent 3° and 5°-wide notches respectively, plotted for visual reference. The horizontal line on the typical error reflects an 8.6° angular averaging within the 48-s data interval produced by the spacecraft as it rolls in clock space, while the vertical line reflects the statistical uncertainty in the number of counts. Count reductions appear broadened in both pitch angle and clock angle space, reflecting the ~40° opening angle of the telescope.

Next, we calculate a 70°-offset rate reduction ($\delta_{70°}$) from the roll maneuver fit results by summing the counts in HET 1's simulated bi-directional response function (Figure 17b) with and without the notch cut. Figure 17a shows HET 1's observed count rate during 70°-offsets on days 2015-297 through 2015-299, a subset of the full sequence of offset maneuvers which took place nearest to the 2015-310 roll (see Figure 3). The observed 70°-offset reduction for this interval was 11.0% ± 0.3% and the simulated value was 12.2% ± 1.2%. Figure 18 shows the same concept applied to HET 2.

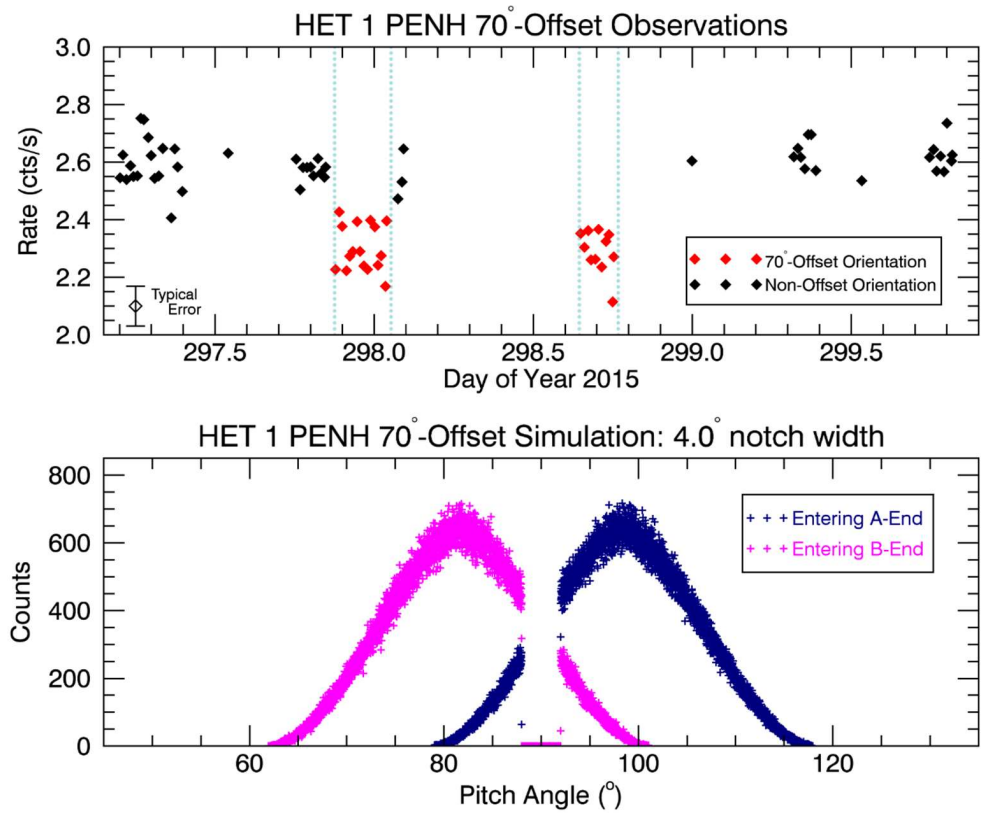

**Figure 17.** HET 1's 70°-offset observed intensities (a) and simulation (b) near the time of the 2015-310 roll maneuver.
(a) An average (typically 480-s intervals) of a subset of data from the offset sequence which began on 2015-296. The full set of maneuvers consisted of 7 offsets that took place between days 296 and 312 (see Figure 3) and the observed rate reduction was 11.0% ± 0.3% for this series of maneuvers.
(b) The 4.0°-wide notch cut applied to a model of HET 1's bi-directional 70°-offset response function for particles entering the telescope's A-end (right; navy blue) and B-end (left; pink). The simulated reduction was 12.2% ± 1.2%.

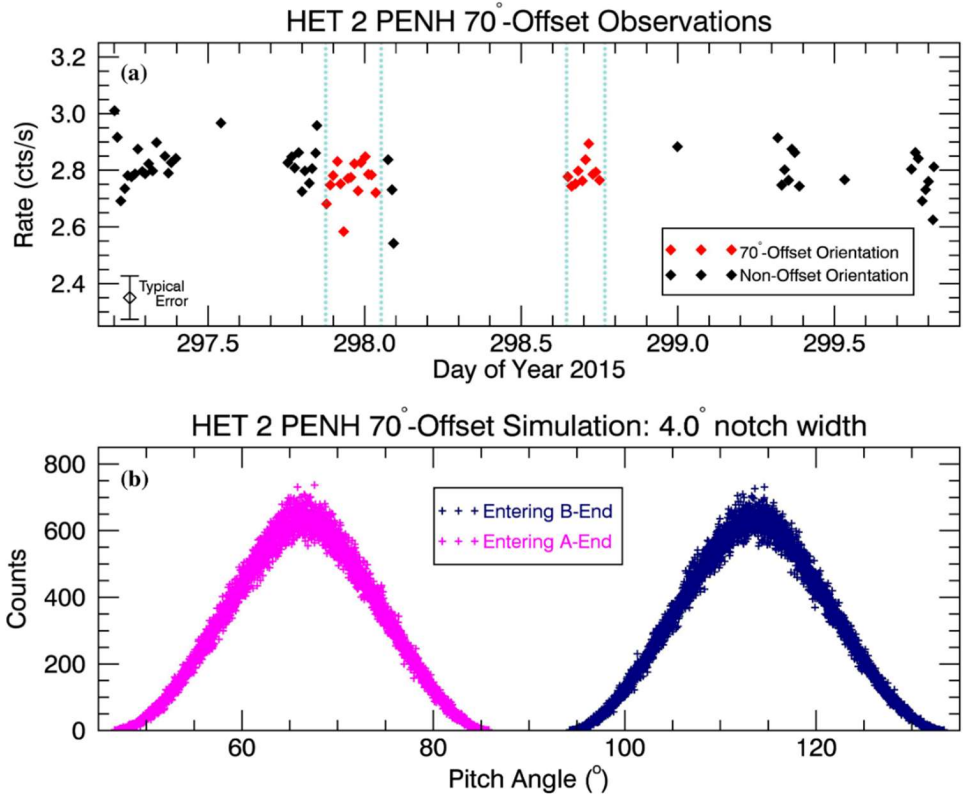

**Figure 18.** Similar to Figure 17, but for HET 2.
(a) HET 2's observed rate reduction was 1.7% ± 0.4% for this series of maneuvers.
(b) Due to HET 2's boresight orientation, Model #1 produces no reduction when a 4.0°-wide notch is applied to this telescope's simulated bi-directional 70°-offset response function. Note that HET 1 and HET 2 are oppositely oriented, so particles entering the HET 2's B-end are on the right (navy blue) and while particles entering HET 2's A-end are on the left (pink).

Finally, we determine the omnidirectional intensity reduction ($\delta_{omni}$) by applying the best-fit roll maneuver notch to Equation A3. For example, 2015-310's 4.0° ± 0.4° effective notch width produces a 3.5% ± 0.3% reduction, which is consistent with the observed value: 3.3% ± 0.1% (comparable to the rates near ~2015.8 and ~2013.85 in Figure 1b).

**A.3 Monte Carlo Simulation Procedure**
We numerically simulate particles passing through HET 1 and HET 2 using the procedure outlined below. The directional observations throughout this work use each telescope's PENH rates, for which particles pass entirely through the detector stack. Since this coincidence mode reflects an integrated rate and does not discriminate amongst the various ions, we perform our particle selection in a manner that is independent of particle energy or species. To inform the telescope geometry, we use detectors B1[3] and C1(with radii of 1.596 and 1.739 cm respectively) spaced at a distance of $L = 9.094$ cm (measured from the top of one detector to the bottom of the other). See Stone et al. (1977) for more details about CRS telescopes.

---

[3] The B detectors are curved, thin detectors. In the Monte Carlo simulation, B1 is modeled as flat and its spacing is defined relative to the bottom of its curvature. Although the curvature alters the path length of the particles and can affect the total energy loss, this has negligible effects on the integrated rates. Moreover, treating B1 as flat does not significantly alter a telescope's field of view.

1. Generate a particle on the first detector at a location uniformly randomly distributed in $x_1, y_1$.
2. Generate a random direction for the particle using a $cos^2(\theta)$ distribution.
3. Use these values to calculate the projected points in $x$ and y when a particle travels a distance $L$ in $\hat{z}$.
4. Keep only the projected points which pass through both detectors. Label these particle coordinates – defined with respect to the top detector – as $(p_x, p_y, p_z)$.
5. Convert particle coordinates to R, T, N: $(p_x, p_y, p_z) \rightarrow (p_r, p_t, p_n)$.
6. Calculate the pitch angle by taking the dot product between the particle's coordinates and the observed B-field direction for a particular maneuver interval (in R, T, N).
7. For a given magnetic field direction and telescope viewing direction (different orientations for HET 1 and HET 2, for example), output information about the telescope orientation (clock angle, θ) and particle pitch angles (α).
8. Simulate a roll maneuver by rotating the spacecraft about $\hat{R}$ in small clock angle increments over 360° (in R, T, N) and repeat steps 1-7 to accumulate the desired number of particles.
9. Simulate the 70°-offset data by fixing clock angle at 70° – roughly a 70°-offset rotation about $\hat{R}$ (in R, T, N) – and repeat steps 1-7 to accumulate the desired number of particles.

## A.4 Model #1 Results

Here, we summarize Model #1's results in tabular form. Table 3 lists the effective notch widths, simulated and observed 70°-offset intensity reductions, and simulated and observed omnidirectional intensity reductions – all for HET 1 (also plotted in Figure 4). Table 4 lists the same quantities for HET 2 (plotted in Figure 5).

| Roll Maneuver Interval | Effective Notch Width | 70°-offset Simulations (HET 1 PENH) | 70°-offset Observations (HET 1 PENH) | Omnidirectional Simulations (HET 1 Guards) | Omnidirectional Observations (HET 1 Guards) |
|---|---|---|---|---|---|
| 2012-263 | 2.5° ± 0.4° | 0.991 ± 0.002 | NA | 0.978 ± 0.003 | 0.995 ± 0.002 |
| 2012-307 | 2.6° ± 0.5° | 0.959 ± 0.007 | 0.973 ± 0.004 | 0.977 ± 0.004 | 0.989 ± 0.005 |
| 2013-31 | 1.1° ± 0.4° | 0.986 ± 0.005 | NA | 0.990 ± 0.003 | 0.995 ± 0.001 |
| 2013-71 | 2.0° ± 0.4° | 0.963 ± 0.007 | 0.957 ± 0.004 | 0.982 ± 0.003 | 0.984 ± 0.001 |
| 2013-122 | 3.7° ± 0.4° | 0.925 ± 0.008 | 0.939 ± 0.004 | 0.967 ± 0.003 | 0.976 ± 0.001 |
| 2013-214 | 0.9° ± 0.4° | 0.983 ± 0.008 | 0.991 ± 0.004 | 0.992 ± 0.003 | 0.996 ± 0.001 |
| 2013-261 | 0.1° ± 0.3° | 0.997 ± 0.006 | NA | 0.999 ± 0.002 | 1.001 ± 0.001 |
| 2013-305 | 0.4° ± 0.5° | 0.992 ± 0.008 | 0.994 ± 0.004 | 0.996 ± 0.004 | 0.999 ± 0.001 |
| 2014-30 | 0.3° ± 0.4° | 0.995 ± 0.006 | 0.994 ± 0.004 | 0.998 ± 0.003 | 0.998 ± 0.001 |
| 2014-69 | 0.3° ± 0.4° | 0.992 ± 0.009 | NA | 0.997 ± 0.003 | 0.995 ± 0.001 |
| 2014-121 | 1.2° ± 0.4° | 0.976 ± 0.008 | 0.989 ± 0.004 | 0.989 ± 0.003 | 1.006 ± 0.001 |
| 2014-213 | 1.6° ± 0.4° | 0.959 ± 0.010 | NA | 0.986 ± 0.003 | 0.993 ± 0.001 |
| 2014-260 | 0.2° ± 0.3° | 0.997 ± 0.004 | 1.012 ± 0.004 | 0.998 ± 0.002 | 0.998 ± 0.001 |
| 2014-304 | 0.8° ± 0.5° | 0.987 ± 0.009 | 1.007 ± 0.005 | 0.993 ± 0.004 | 0.994 ± 0.001 |
| 2015-36 | 0.9° ± 0.5° | 0.990 ± 0.006 | 0.994 ± 0.004 | 0.993 ± 0.004 | 0.991 ± 0.001 |
| 2015-127 | 1.1° ± 0.5° | 0.991 ± 0.004 | 0.980 ± 0.004 | 0.990 ± 0.004 | 0.980 ± 0.001 |
| 2015-219 | 2.5° ± 0.4° | 0.956 ± 0.007 | 0.929 ± 0.004 | 0.978 ± 0.003 | 0.977 ± 0.001 |
| 2015-252 | 3.4° ± 0.4° | 0.945 ± 0.007 | 0.928 ± 0.006 | 0.970 ± 0.003 | 0.975 ± 0.001 |
| **2015-257** | 2.6° ± 0.4° | 0.950 ± 0.008 | 0.928 ± 0.006 | 0.977 ± 0.003 | 0.969 ± 0.001 |
| **2015-310** | 4.0° ± 0.4° | 0.878 ± 0.012 | 0.890 ± 0.003 | 0.965 ± 0.003 | 0.967 ± 0.001 |
| **2016-35** | 3.8° ± 0.6° | 0.909 ± 0.013 | 0.886 ± 0.004 | 0.967 ± 0.005 | 0.981 ± 0.001 |
| **2016-84** | 2.3° ± 0.5° | 0.960 ± 0.008 | NA | 0.980 ± 0.004 | 0.986 ± 0.001 |
| **2016-126** | 1.8° ± 0.5° | 0.996 ± 0.001 | 0.967 ± 0.004 | 0.985 ± 0.004 | 0.985 ± 0.001 |
| **2016-218** | 1.6° ± 0.4° | 0.995 ± 0.001 | 0.951 ± 0.004 | 0.986 ± 0.003 | 0.989 ± 0.001 |
| **2016-309** | 0.7° ± 0.4° | 0.998 ± 0.001 | 0.982 ± 0.009 | 0.994 ± 0.003 | 0.989 ± 0.001 |

**Table 3.** A summary of effective notch widths (obtained from bi-directional roll maneuver fits to PENH rates; ≳70 MeV, proton-dominated) and corresponding relative intensity changes arising from the particle pitch-angle anisotropy for simulated and observed 70°-offset and omnidirectional observations for HET 1. The time periods shown in red indicate intervals during which the anisotropy is most prominent, superimposed in Figure 2. Simulated intensities are normalized to values obtained from notch-free simulated response functions. Observed 70°-offset intensities are normalized to temporally-adjacent non-offset rates and omnidirectional observations are normalized to the average values during the 2013.6 to 2014.1 time period when count rates are relatively uniform and isotropic. Data are plotted in Figure 4.

| Roll Maneuver Interval | Effective Notch Width | 70°-offset Simulations (HET 1 PENH) | 70°-offset Observations (HET 1 PENH) | Omnidirectional Simulations (HET 1 Guards) | Omnidirectional Observations (HET 1 Guards) |
|---|---|---|---|---|---|
| **2012-263** | 1.3° ± 0.5° | 0.993 ± 0.003 | NA | 0.989 ± 0.004 | 0.993 ± 0.002 |
| **2012-307** | 2.4° ± 0.4° | 0.985 ± 0.003 | 0.989 ± 0.003 | 0.979 ± 0.004 | 0.983 ± 0.005 |
| **2013-31** | 1.6° ± 0.5° | 0.999 ± 0.000 | NA | 0.986 ± 0.004 | 0.991 ± 0.002 |
| **2013-71** | 3.0° ± 0.4° | 1.000 ± 0.000 | 0.987 ± 0.004 | 0.974 ± 0.004 | 0.982 ± 0.001 |
| **2013-122** | 3.6° ± 0.4° | 1.000 ± 0.000 | 0.984 ± 0.004 | 0.968 ± 0.004 | 0.973 ± 0.001 |
| **2013-214** | 1.4° ± 0.4° | 1.000 ± 0.000 | 0.999 ± 0.004 | 0.988 ± 0.004 | 0.996 ± 0.001 |
| **2013-261** | 0.0° ± 0.3° | 1.000 ± 0.000 | NA | 1.000 ± 0.002 | 1.001 ± 0.001 |
| **2013-305** | 1.0° ± 0.7° | 1.000 ± 0.000 | 0.996 ± 0.004 | 0.991 ± 0.006 | 1.000 ± 0.001 |
| **2014-30** | 0.5° ± 0.4° | 1.000 ± 0.000 | 0.995 ± 0.004 | 0.996 ± 0.004 | 0.998 ± 0.001 |
| **2014-69** | 1.4° ± 0.5° | 1.000 ± 0.000 | NA | 0.988 ± 0.004 | 0.995 ± 0.001 |
| **2014-121** | 1.6° ± 0.4° | 1.000 ± 0.000 | 1.014 ± 0.004 | 0.986 ± 0.004 | 1.007 ± 0.001 |
| **2014-213** | 0.8° ± 0.6° | 0.998 ± 0.001 | NA | 0.993 ± 0.005 | 0.992 ± 0.001 |
| **2014-260** | 0.1° ± 0.3° | 1.000 ± 0.000 | 0.998 ± 0.004 | 0.999 ± 0.002 | 1.001 ± 0.001 |
| **2014-304** | 0.3° ± 0.5° | 1.000 ± 0.000 | 1.000 ± 0.005 | 0.997 ± 0.004 | 0.997 ± 0.001 |
| **2015-36** | 0.7° ± 0.4° | 1.000 ± 0.000 | 1.001 ± 0.004 | 0.994 ± 0.004 | 0.995 ± 0.001 |
| **2015-127** | 1.4° ± 0.4° | 1.000 ± 0.000 | 0.999 ± 0.004 | 0.987 ± 0.004 | 0.990 ± 0.001 |
| **2015-219** | 2.5° ± 0.4° | 1.000 ± 0.000 | 0.990 ± 0.003 | 0.978 ± 0.004 | 0.980 ± 0.001 |
| **2015-252** | 3.1° ± 0.5° | 1.000 ± 0.000 | 0.985 ± 0.006 | 0.973 ± 0.004 | 0.976 ± 0.001 |
| **2015-257** | 2.8° ± 0.4° | 1.000 ± 0.000 | 0.985 ± 0.006 | 0.976 ± 0.004 | 0.976 ± 0.001 |
| **2015-310** | 3.0° ± 0.4° | 1.000 ± 0.000 | 0.983 ± 0.004 | 0.974 ± 0.004 | 0.970 ± 0.002 |
| **2016-35** | 3.9° ± 0.7° | 1.000 ± 0.000 | 0.990 ± 0.004 | 0.966 ± 0.006 | 0.970 ± 0.001 |
| **2016-84** | 2.0° ± 0.5° | 1.000 ± 0.000 | NA | 0.983 ± 0.004 | 0.982 ± 0.001 |
| **2016-126** | 0.9° ± 0.6° | 1.000 ± 0.000 | 0.998 ± 0.004 | 0.992 ± 0.005 | 0.987 ± 0.001 |
| **2016-218** | 1.7° ± 0.5° | 1.000 ± 0.000 | 0.999 ± 0.004 | 0.986 ± 0.004 | 0.981 ± 0.001 |
| **2016-309** | 0.0° ± 0.3° | 1.000 ± 0.000 | 1.011 ± 0.008 | 1.000 ± 0.002 | 0.987 ± 0.001 |

**Table 4.** Similar to Table 3, but for HET 2 (plotted in Figure 5).

# Appendix B – Model #2: Partially-Filled Notch
## B.1 Omnidirectional Response Function
Model #2 utilizes a two-parameter representation of the notch by introducing a depth term to allow for the possibility of scattering. We achieve this in the omnidirectional notch response function by modifying $j_n$ in Equation A2 to allow for a reduced directional intensity representation of the missing particle distribution ($j_n < j_0$), leading to:

$$\delta_{omni} = \frac{J_n}{J} = \frac{j_n}{j_0} \cos(90° - w/2) = d \times \mu. \quad (B1)$$

Hence, the notch is now partially filled and characterized by an "effective area" of depth, $d = \frac{j_n}{j_0}$ and width $\mu = \cos \alpha$ ranging from $\alpha = 90° + w/2$ to $\alpha = 90° - w/2$, as shown in Figure 19.

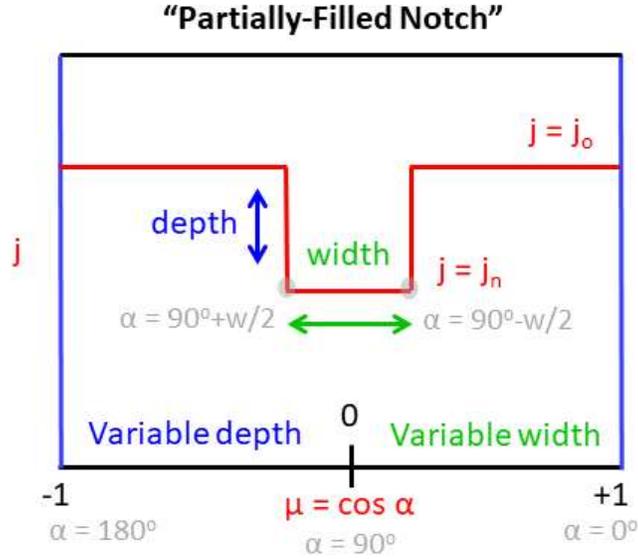

**Figure 19.** Diagram of notch model #2.

For a given period where the anisotropy is prominent ($\delta_{omni} > 0$), the range of possible widths is no larger than LECP's full telescope opening angle: $0° \leq w \leq 45°$. Therefore, the range of possible depths is given by:

$$d = \frac{\delta_{omni}}{\cos(90° - w/2)}. \quad (B2)$$

**B.2 Directional Response Functions**

Since the anisotropy is now represented as a single function with two unknowns – width and depth– we employ alternate strategies to implement and evaluate the effectiveness of Model #2's simulations. One strategy is to extend the roll maneuver fits to allow for two parameters – variable width and depth. We thereby acquire independent best-fit notch geometries for HET 1 and HET 2 and compare their results. For example, the 120-122 roll maneuver is characterized by a nominal width of 25.8° (ranging from 23.2° to 33.4°) and depth of 18.5% (ranging from 21.5% to 15.5%) for HET 1. For HET 2, the nominal width and depth is 24.6° and 18.0% (ranging from 20.3° to 29.2° and 21.0% to 15.0%, respectively).

A second strategy is to determine widths and depths that achieve consistency between each telescope's omnidirectional and 70°-offset response functions, as detailed in the following subsection. The assumption here is that omnidirectional and directional rates are responding to the same notch geometry and the expectation is that response function curves should differ enough to set at least some limits on the notch's widths and depths.

## B.2.1 Omnidirectional and 70°-offset Response Function Curves

Due to the time-varying nature of the anisotropy and a weak intensity reduction observed by HET 2 during its 70°-offsets, the analysis for Model #2 focuses on the 6 offset intervals where the anisotropy is most prominent, listed in Table 5. The telescope orientations, omnidirectional intensity reductions ($\delta_{omni}$), 70°-offset reductions ($\delta_{70°}$), and magnetic field observations all reflect the average taken over the offset maneuver sequence time periods.

| Offset Interval | 2013-67 | 2013-120 | 2015-208 | 2015-250 | 2015-296 | 2016-31 |
|---|---|---|---|---|---|---|
| Maneuver Days | 67, 68, 69, 70, 71 | 120, 121, 122 | 208, 209, 210, 215, 216 | 250, 251, 252 | 296, 297, 298, 300, 301, 302, 303, 307, 308, 312 | 31, 32, 34, 38, 39, 40 |
| HET 1 Offset Boresight (A-end) | R = -0.494<br>T = -0.675<br>N = -0.548 | -0.495<br>-0.673<br>-0.550 | -0.503<br>-0.669<br>-0.547 | -0.506<br>-0.669<br>-0.545 | -0.505<br>-0.671<br>-0.543 | -0.496<br>-0.678<br>-0.543 |
| HET 1 Boresight Pitch Angle | α = 78.5° | 79.3° | 77.2° | 76.7° | 81.3° | 78.3° |
| HET 2 Offset Boresight (B-end) | R = -0.209<br>T = -0.056<br>N = 0.976 | -0.212<br>-0.056<br>0.976 | -0.210<br>-0.051<br>0.976 | -0.207<br>-0.051<br>0.977 | -0.204<br>-0.050<br>0.978 | -0.206<br>-0.050<br>0.977 |
| HET 2 Boresight Pitch Angle | α = 69.2° | 70.0° | 66.1° | 67.1° | 66.2° | 67.2° |
| 70°-offset Redution ($\delta_{70°}$) | HET 1 = 4.3 ± 0.4%<br>HET 2 = 1.3 ± 0.4% | 6.1 ± 0.4%<br>1.6 ± 0.4% | 7.1 ± 0.4%<br>1.0 ± 0.4% | 7.2 ± 0.6%<br>1.5 ± 0.6% | 11.0 ± 0.3%<br>1.7 ± 0.4% | 11.4 ± 0.4%<br>1.0 ± 0.4% |
| Omnidirectional Reduction ($\delta_{omni}$) | HET 1 = 1.6 ± 0.05%<br>HET 2 = 1.8 ± 0.05% | 2.4 ± 0.05%<br>2.7 ± 0.05% | 1.9 ± 0.05%<br>2.1 ± 0.05 % | 2.2 ± 0.07%<br>2.4 ± 0.07% | 2.9 ± 0.04%<br>3.1 ± 0.05% | 3.1 ± 0.06%<br>3.3 ± 0.06% |
| B field (nT) | R = 0.175<br>T = -0.444<br>N = 0.200<br>\|B\| = 0.517 | 0.178<br>-0.421<br>0.188<br>0.495 | 0.118<br>-0.402<br>0.197<br>0.463 | 0.117<br>-0.392<br>0.183<br>0.448 | 0.152<br>-0.379<br>0.200<br>0.455 | 0.132<br>-0.370<br>0.180<br>0.433 |

**Table 5.** Summary of HET 1 and HET 2 observational values used for 70°-offset and omnidirectional simulations.

From the observed omnidirectional intensity reduction ($\delta_{omni}$), we determine the combination of widths and depths that produce results consistent with the observations and uncertainties using widths ($0° \leq w \leq 45°$) and depths informed by Equation B2. We take a similar approach to determine the 70°-offset isocontours, using the observed relative intensity reduction ($\delta_{70°}$) and a given telescope's simulated response function to numerically determine depth (analogous to solving Equation B2).

Figures 20 & 21 show superimposed omnidirectional and 70°-offset curves for the 2013-120 offset. While HET 1's observations allow for a broad range of widths and depths – 2.8° to > 45° and 100% to < 6.5% (Figure 20) – HET 2's observations narrow the range of possible values to widths of 19.2° to 25.8° and respective depths of 16.4% down to 11.8% along the curve (Figure 21). The 2013-120 nominal values are 22.5° wide and 13.7% deep.

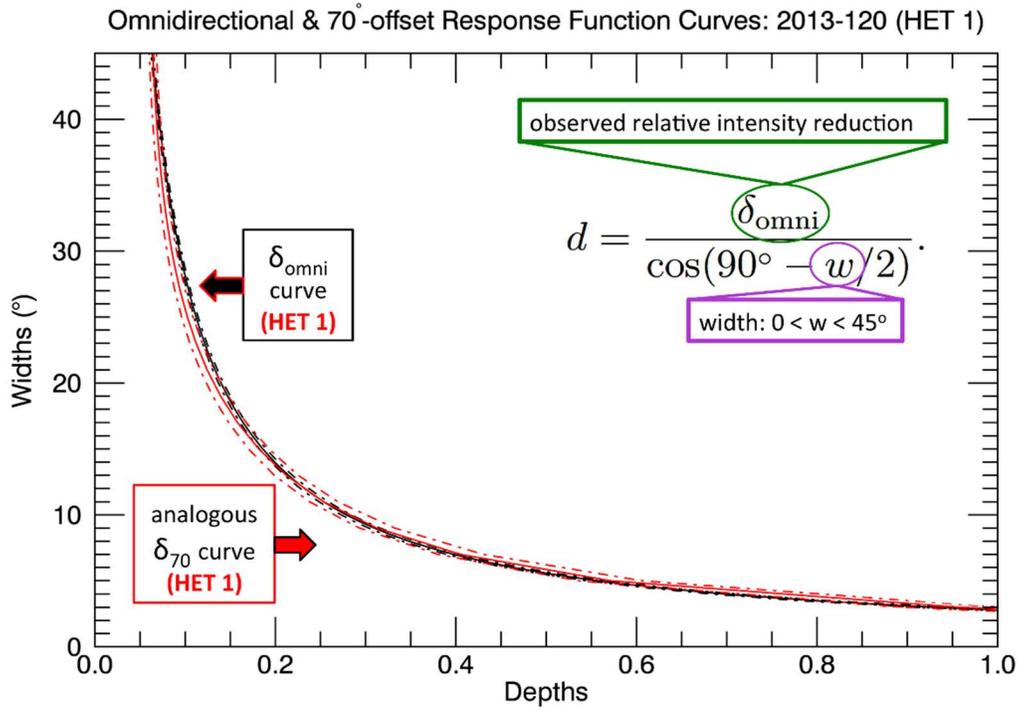

**Figure 20.** Simulated omnidirectional (black, solid) and 70°-offset (solid, red) widths vs. depths for HET 1 during the 2013-120 offset. The dotted curves reflect the 1-σ uncertainties in the omnidirectional (black) and 70°-offset (red) measurements.

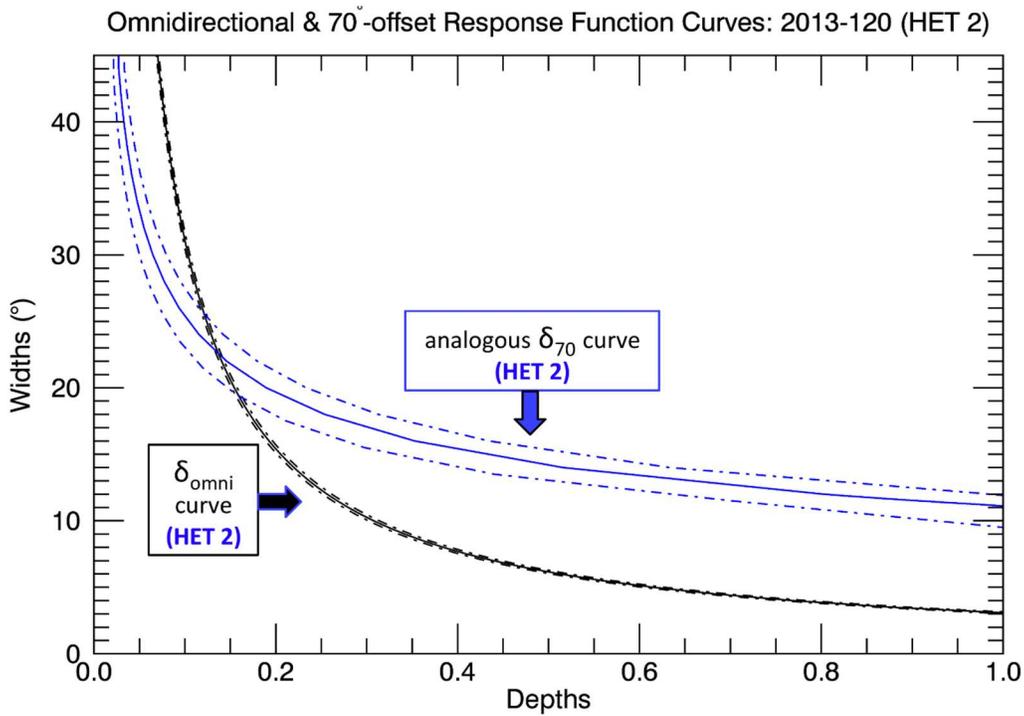

**Figure 21.** HET 2 version of Figure 20.

Differences in HET 1 & 2's boresight orientations enable HET 2 to better set limits to the notch's parameters than HET 1. Indeed, this is true for all offsets. HET 2 is most sensitive to the notch's width and depth since the anisotropy is at the very edge of its field of view. In contrast, the anisotropy is more fully within HET 1's field of view, so it is much more sensitive to the magnetic field direction than HET 2.

To illustrate each telescope's differing sensitivities, HET 1's width vs. depth curves for boresight pitch angles ranging from α = 70° to 85° are presented for the 2013-120 offset in Figure 22. Similar curves are also shown for HET 2, in this case for boresight pitch angles ranging from α = 60° to 75° (Figure 23). Notably, each of HET 2's simulated 70°-offset curves (solid blue) intersect with the omnidirectional curves (dashed black) at some point, revealing a variety of possible solutions, depending on the particular value of α. However, the majority of HET 1's curves do not intersect; the few which do represent a narrow range of pitch angles, with curves overlapping so well that the range of possible widths and depths is not effectively constrained by HET 1 alone.

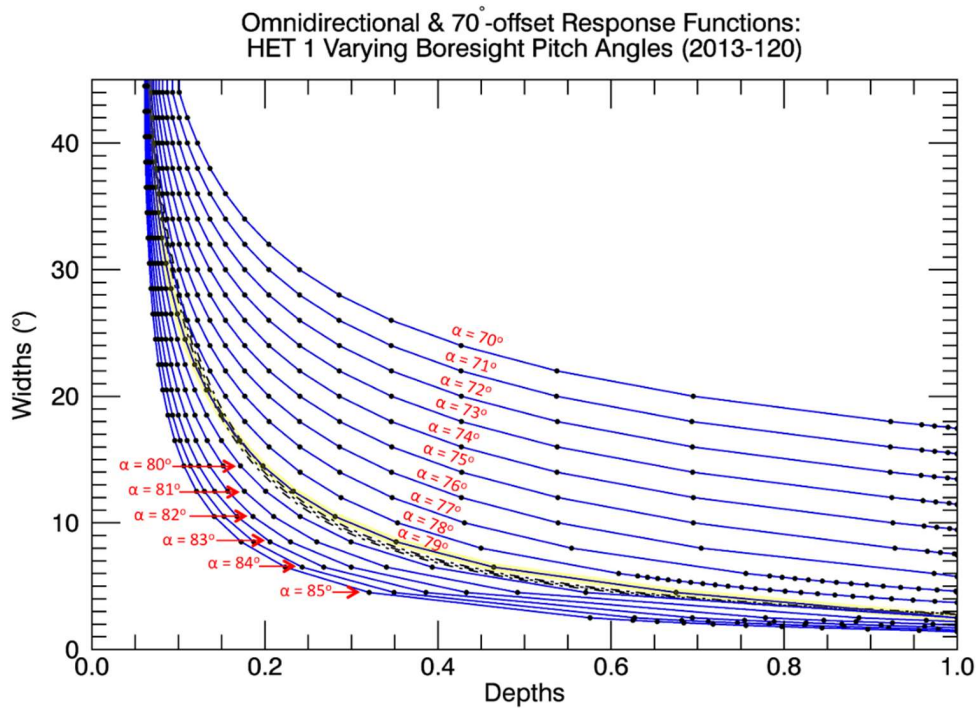

**Figure 22.** Width and depth curves for simulated HET 1 (PENH; ≳70 MeV, proton-dominated) 70°-offset (blue, solid) and omnidirectional (black, dotted) notch response functions for pitch angles ranging from α = 70° to 85°. These pitch angles reflect the angle between the telescope's B-end boresight with respect to the magnetic field. The 70°-offset curves were each calculated from observations listed in Table 5 (uncertainties not shown). HET 1's boresight pitch angle during the 2013-120 sequence of 70°-offsets was α = 79.3° (yellow).

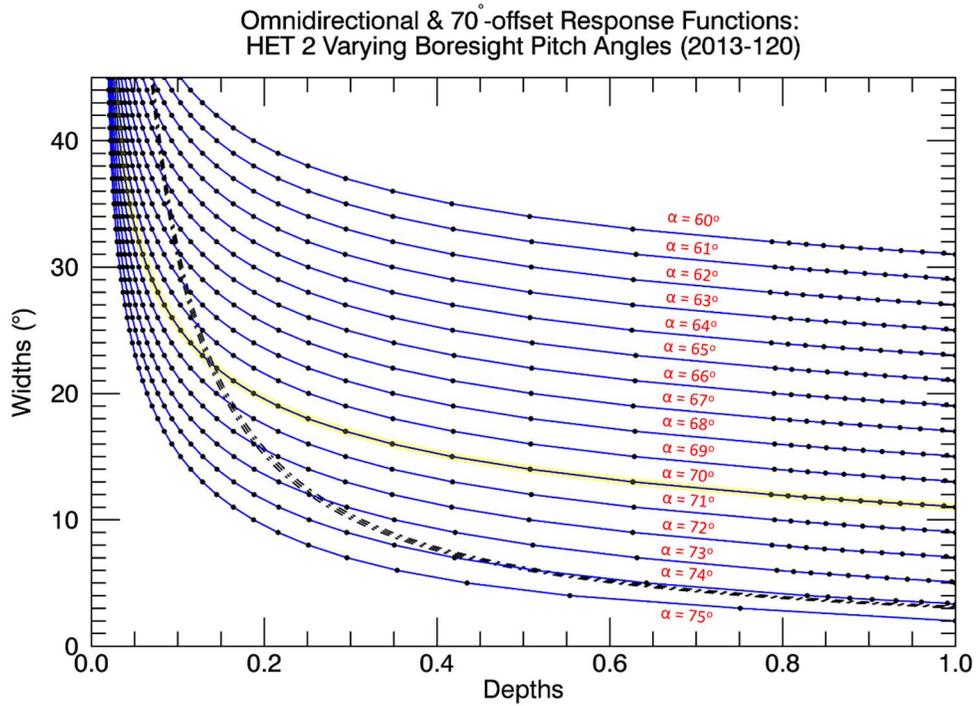

**Figure 23.** Width and depth curves for simulated HET 2 (PENH; ≳ 70 MeV, proton-dominated) 70°-offset (blue, solid) and omnidirectional (black, dotted) notch response functions, for pitch angles ranging from α = 60° to 75°. The pitch angles shown are with respect to HET 2's A-end boresight; its nominal 70°-offset boresight pitch angle during the 2013-120 sequence was α = 70.0° (yellow).

### B.2.2 Model #2's Alternative Magnetic Fields

A complication arises because the combination of the telescope's boresight direction and the observed magnetic field direction for most intervals produces disagreement between HET 1's omnidirectional and 70°-offset notch response function curves. In fact, the 2013-120 is the only interval where omnidirectional and 70°-offset simulations intersect without adjustment (albeit, only a small adjustment is needed for 2013-67). For example, Figure 24 shows HET 1's curves for 2016-31. For this interval there is no strong agreement between the two curves to within their respective uncertainties that also yields a width and depth consistent with HET 2 observations.

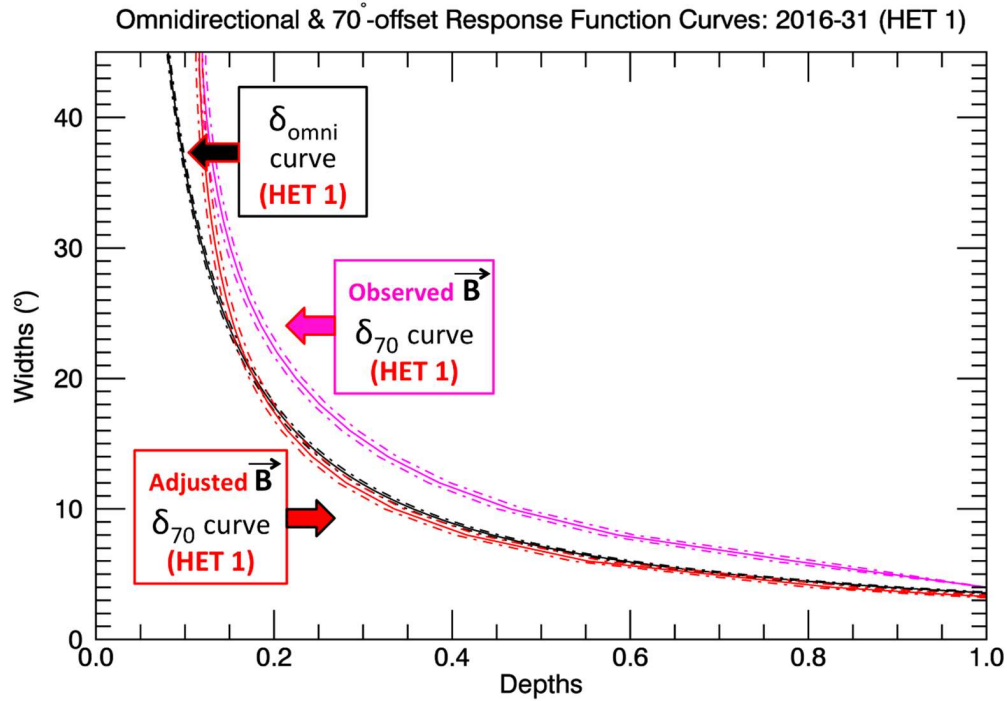

**Figure 24.** Simulated HET 1 omnidirectional (black, dotted) and two 70°-offset response function curves indicating the difference between the observed B-field (pink) of (0.132, -0.370, 0.180) nT (in R, T, N) and an adjusted B-field (red) of (0.181, -0.351, 0.170) nT during the 2016-31 offset. HET 1's B-end 70°-offset boresight pitch angle was 79.3° for the observed case and 82.5° for the adjusted case.

Ultimately, the typical shift in boresight pitch angle required to resolve HET 1's disagreement (~3.5° in α) is larger than CRS's expected telescope alignment uncertainties (≲ 1°). However, agreement between HET 1 and HET 2 can be achieved using a B-field that falls within the magnetometer's 1-σ uncertainties, δB = (±0.06, ±0.02, ±0.02) nT in R, T, N. Thus, we perform an additional search for alternate B-field directions that achieve agreement amongst HET 1 & 2 omnidirectional and directional observations for each interval. The results of this search are listed in Table 6 and shown in Figure 25. In principle, differing combinations of $B_r$, $B_t$, and $B_n$ can produce identical pitch angles for HET 1, so other solutions could exist. Nonetheless, we select each component by minimizing its deviation from the reported measurement.

| | | 2013-67 | 2013-120 | 2015-208 | 2015-250 | 2015-296 | 2016-31 |
|---|---|---|---|---|---|---|---|
| **Observed Magnetic Field (nT)** | Br = | 0.175 | 0.178 | 0.118 | 0.117 | 0.152 | 0.132 |
| | Bt = | -0.444 | -0.421 | -0.402 | -0.392 | -0.379 | -0.370 |
| | Bn = | 0.200 | 0.188 | 0.197 | 0.183 | 0.200 | 0.180 |
| | \|B\| = | 0.517 | 0.495 | 0.463 | 0.448 | 0.455 | 0.433 |
| **Alternative Magnetic Field (nT)** | Br = | 0.180 | 0.178 | 0.178 | 0.173 | 0.196 | 0.181 |
| | Bt = | -0.440 | -0.421 | -0.382 | -0.372 | -0.384 | -0.351 |
| | Bn = | 0.207 | 0.188 | 0.207 | 0.169 | 0.180 | 0.170 |
| | \|B\| = | 0.519 | 0.494 | 0.470 | 0.444 | 0.467 | 0.430 |
| **ΔB (nT)** | ΔBr = | 0.005 | 0.000 | 0.060 | 0.056 | 0.044 | 0.049 |
| | ΔBt = | 0.004 | 0.000 | 0.020 | 0.020 | -0.005 | 0.019 |
| | ΔBn = | 0.007 | 0.000 | 0.010 | -0.014 | -0.020 | -0.010 |

**Table 6.** Summary of observed and predicted magnetic fields used for the variable width, variable depth notch analysis. ΔB represents the difference between the alternative and observed magnetic fields. The magnetometer's 1-σ uncertainties are δB = (±0.06, ±0.02, ±0.02) nT nT in R, T, N.

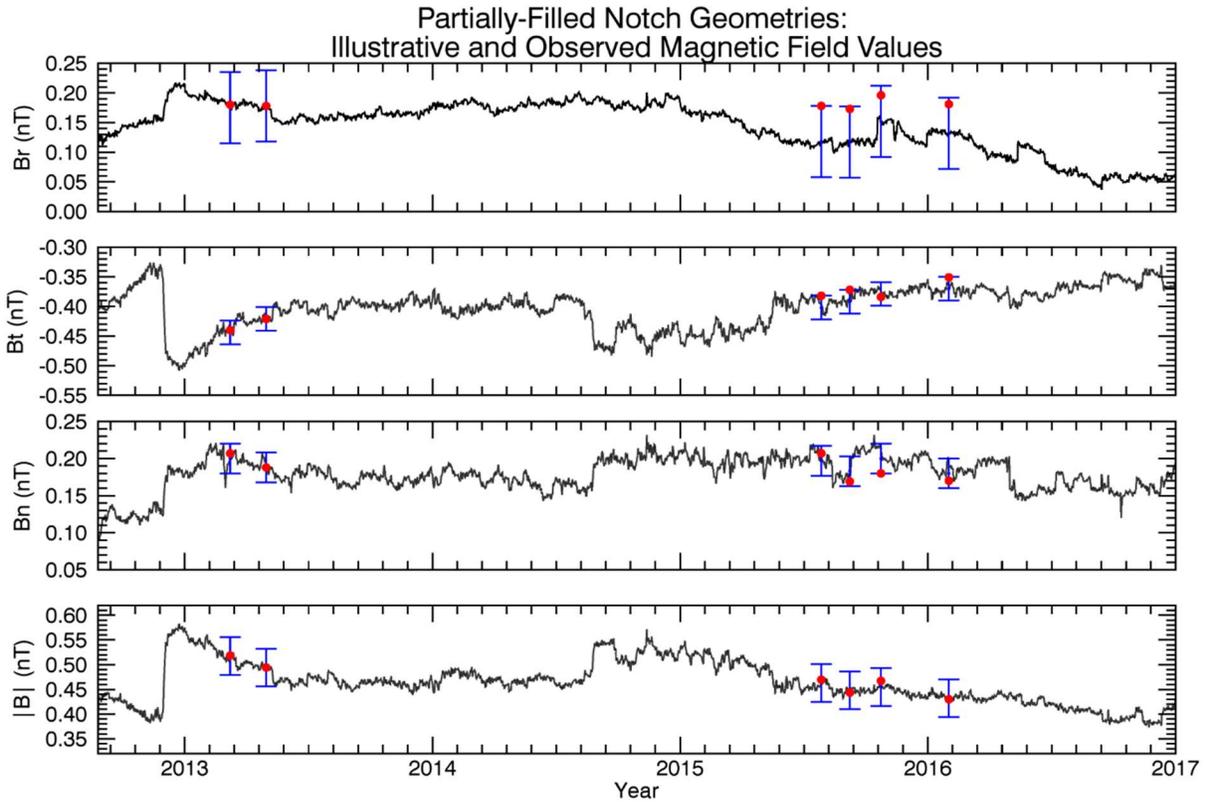

**Figure 25.** Comparison of observed and alternative magnetic fields used for Model #2's variable width, variable depth notch analysis. The error bars (blue) reflect the magnetometer's 1-σ uncertainties: δB = (±0.06, ±0.02, ±0.02) nT in R, T, N.

### B.3 Model #2 Results
Tables 7 & 8 list the fit results for independently calculated roll maneuvers for HET 1 and HET 2, allowing for notches of variable width and depth. Tables 9 & 10 list the notch parameters obtained using 70°-offset and omnidirectional response function curves for HET 1 and HET 2, respectively.

| HET 1: Interval | Nominal Width | Lower Limit | Upper Limit | Nominal Depth | Upper Limit | Lower Limit | P-Value of $\chi^2$ Fit |
|---|---|---|---|---|---|---|---|
| 2013-71  | 26.8° | 19.0° | 35.9° | 9.4%  | 12.4% | 6.4%  | 62.5% |
| 2013-122 | 25.8° | 23.2° | 33.4° | 18.5% | 21.5% | 15.5% | 48.7% |
| 2015-219 | 28.8° | 22.5° | 34.9° | 12.3% | 16.3% | 9.3%  | 94.9% |
| 2015-252 | 25.7° | 21.2° | 30.7° | 15.8% | 18.8% | 12.8% | 58.5% |
| 2015-310 | 20.8° | 17.9° | 25.0° | 22.4% | 27.4% | 18.4% | 18.5% |
| 2016-35  | 13.4° | 10.6° | 16.4° | 29.5% | 36.5% | 23.5% | 21.8% |

**Table 7.** HET 1 roll maneuver fits for notches of variable width and depth for the 6 intervals of Model #2. P-values for all intervals are >5%, indicative of good $\chi^2$ fits. Results are plotted in Figures 8 & 9.

| HET 2: Interval | Nominal Width | Lower Limit | Upper Limit | Nominal Depth | Upper Limit | Lower Limit | P-Value of $\chi^2$ Fit |
|---|---|---|---|---|---|---|---|
| 2013-71 | 34.3° | 29.1° | 42.3° | 12.6% | 16.6% | 9.6% | 56.6% |
| 2013-122 | 24.6° | 20.3° | 29.2° | 18.0% | 21.0% | 15.0% | 70.0% |
| 2015-219 | 17.6° | 13.2° | 22.6° | 15.1% | 19.1% | 11.1% | 14.0% |
| 2015-252 | 10.6° | 8.4° | 13.1° | 28.9% | 34.9% | 22.9% | 0.5% |
| 2015-310 | 20.8° | 17.1° | 28.8° | 16.1% | 19.1% | 13.1% | 30.9% |
| 2016-35 | 15.2° | 11.7° | 18.9° | 26.6% | 33.6% | 20.6% | 70.7% |

**Table 8.** Similar to Table 7, but for HET 2. The P-values for 5 out of 6 intervals are >5%, indicative of good $\chi^2$ fits. The exception occurs during the 2015-252 interval, which is proximate to a plasma oscillation that began on ~2015-247. Results are plotted in Figure 8.

| HET 1: Interval | Range of Widths | Range of Depths |
|---|---|---|
| 2013-67 | 2.1° to > 45° | 100% to < 4.1% |
| 2013-120 | 2.8° to > 45° | 100% to < 6.5% |
| 2015-208 | 11.1° to 26.5° | 18.6% to 8.3% |
| 2015-250 | 2.4° to 33.7° | 100% to 7.7% |
| 2015-296 | 3.3° to 20.6° | 100% to 16.5% |
| 2016-31 | 13.5° to 25.0° | 26.0% to 14.6% |

**Table 9.** HET 1 range of widths and depths from intersection of omnidirectional and 70°-offset response function curves for the 6 intervals where the anisotropy is most prominent. The simulations incorporated values listed in Table 5 and pitch angles determined by the alternative B-fields in Table 6. These results are plotted in Figure 8.

| HET 2: Interval | Nominal Width | Lower Limit | Upper Limit | Nominal Depth | Upper Limit | Lower Limit |
|---|---|---|---|---|---|---|
| 2013-67 | 29.2° | 24.2° | 34.3° | 7.0% | 8.6% | 5.8% |
| 2013-120 | 22.5° | 19.2° | 25.8° | 13.7% | 16.4% | 11.8% |
| 2015-208 | 28.1° | 23.6° | 32.2° | 8.7% | 10.6% | 7.4% |
| 2015-250 | 20.4° | 14.4° | 26.2° | 13.4% | 19.5% | 10.2% |
| 2015-296 | 18.1° | 15.4° | 20.8° | 19.8% | 23.6% | 17.0% |
| 2016-31 | 14.3° | 10.5° | 17.6° | 26.3% | 36.2% | 20.9% |

**Table 10.** HET 2 nominal widths and depths (with ranges) from intersection of omnidirectional and 70°-offset response function curves incorporating values listed in Table 5 and pitch angles determined by the alternative B-fields in Table 6. These results are plotted Figures 8 & 9.